\newif\ifsingle
\singletrue % comment out for single column version

\ifsingle
\documentclass[12pt,draftclsnofoot, onecolumn]{IEEEtran}	
\usepackage[a4paper, total={6.5in, 9.5in}]{geometry}	
\else		
\documentclass[10pt,final, twocolumn]{IEEEtran}
\usepackage[a4paper, total={7.3in, 10.4in}]{geometry}
\fi

\usepackage{times}
\usepackage{amsmath,dsfont}
\usepackage{amssymb,amsthm}
\usepackage{epsfig,verbatim}
\usepackage{subfigure}
\usepackage{setspace}
\usepackage{color}
\usepackage{cite}
\usepackage{epstopdf}
\usepackage{graphics}
\usepackage{accents}
\usepackage{acronym}
\usepackage[bookmarks,colorlinks]{hyperref}
\usepackage{booktabs}
\usepackage{mathtools}
\usepackage{bbm}

\definecolor{NewColor}{rgb}{0,0,0}

\newtheorem{theorem}{Theorem}
\newtheorem{corollary}{Corollary}
\newtheorem{proposition}{Proposition}
\newtheorem{lemma}{Lemma}

\newcommand{\myVec}[1]{{\bf #1}}
\newcommand{\myMat}[1]{{\bf #1}}

\newcommand{\myDetVec}[1]{\myVec{\lowercase{#1}}}
\newcommand{\myRandVec}[1]{\myVec{\lowercase{#1}}}
\newcommand{\myDetMat}[1]{\myMat{\uppercase{#1}}}
\newcommand{\myRandMat}[1]{\myMat{\uppercase{#1}}}
\newcommand{\mySet}[1]{\mathcal{#1}}

\newcommand{\E}{\mathcal{E}}		 			% Stochastic expectation
\newcommand{\myW}{{\myRandVec{W}}}			 		% noise
\newcommand{\myY}{{\myRandVec{Y}}}			 		% observations
\newcommand{\myX}{{\myRandVec{X}}}			 		% Transmitted symbols
\newcommand{\myV}{{\myRandVec{V}}}			 		% Transmitted symbols
\newcommand{\myS}{{\myDetVec{s}}}			 		% Pilot symbols
\newcommand{\myI}{{\myDetMat{i}}}			 		% Identity matrix
\newcommand{\myYmat}{{\myRandMat{Y}}}			 	% observations in matrix form
\newcommand{\mySmat}{{\myDetMat{S}}}			 	% pilots in matrix form
\newcommand{\myWmat}{{\myRandMat{W}}}			 	% noise in matrix form
\newcommand{\Gmat}{{\myRandMat{G}}}			 		% Channel matrix
\newcommand{\Hmat}{{\myRandMat{H}}}			 		% Channel matrix - random part
\newcommand{\Dmat}{{\myDetMat{d}}}			 		% Channel gains
\newcommand{\Bmat}{{\myDetMat{b}}}			 		% Channel gains
\newcommand{\DmatRel}{\bar{\myDetMat{d}}}			 		% Channel gains
\newcommand{\BmatRel}{\bar{\myDetMat{b}}}			 		% Channel gains
\newcommand{\AggMat}{{\myRandMat{Q}}}			 	% observations in matrix form
\newcommand{\Smat}{\tilde{\myDetMat{s}}}			% Pilot symbols matrix
\newcommand{\SigW}{\sigma_W^2}						% Noise variance
\newcommand{\AntRatio}{\kappa}						% Antenna ratio
\newcommand{\Ncells}{n_c}							% Number of cells
\newcommand{\Nantennas}{n_t}						% Number of BS antennas
\newcommand{\Nusers}{n_u}							% Number of users in cell
\newcommand{\NcellsSet}{\mySet{N}_c}				% Number of users in cell
\newcommand{\NusersSet}{\mySet{N}_u}				% Number of users in cell
\newcommand{\dcoeff}{D}								% attenuation coefficient
\newcommand{\dcoeffRel}{d}								% attenuation coefficient
\newcommand{\bcoeff}{B}								% attenuation coefficient
\newcommand{\Tpilots}{\tau_p}						% pilot symbols
\newcommand{\Tdata}{\tau_d}							% data symbols
\newcommand{\Tcoh}{\tau_c}							% coherence duration
\newcommand{\MPFunc}{\nu}								% Marchenko Pastur function
\newcommand{\CovMat}[1]{\myDetMat{c}_{#1}}			% covariance matrix
\newcommand{\UHmat}{ \myMat{M}}
\newcommand{\Xrv}{X}
\newcommand{\Yrv}{Y}
\newcommand{\Xmin}{x_{\min}}
\newcommand{\Xmax}{x_{\max}}
\newcommand{\Ymin}{y_{\min}}
\newcommand{\Ymax}{y_{\max}}
\newcommand{\mxy}{\mu_{\Xrv,\Yrv}}

\newcommand{\IAN}[1]{^{\rm IAN}_{#1}}
\newcommand{\SD}[1]{^{\rm SD}_{#1}}
\newcommand{\TD}[1]{^{\rm TD}_{#1}}
\newcommand{\OS}[1]{^{\rm OS}_{#1}}

\newcommand{\R}[1]{r_{#1}}
\newcommand{\RIAN}[1]{r\IAN{#1}}
\newcommand{\RSD}[1]{r\SD{#1}}
\newcommand{\RTD}[1]{r\TD{#1}}
\newcommand{\ROS}[1]{r\OS{#1}}
\newcommand{\RSEP}[1]{r^{\rm SEP}_{#1}}

\newcommand{\Dist}{ \stackrel{d}{=}}
\newcommand{\AsConv}{\mathop{\longrightarrow}\limits^{\rm a.s.}}
\newcommand{\CDF}[1]{F_{#1}}
\newcommand{\myEta}{\eta}
\newcommand{\myZeta}{\zeta}
\newcommand{\myT}{T}
\newcommand{\myProb}{p}

\newcommand{\AmatUp}{\AggMat^{\rm Net}_{k}}
\newcommand{\AmatDn}{\AggMat^{\rm Int}_{k}}
\newcommand{\TAmatUp}{\tilde{\AggMat}^{\rm Net}_{k}}
\newcommand{\TAmatDn}{\tilde{\AggMat}^{\rm Int}_{k}}
\newcommand{\AmatTD}{\AggMat\TD{k}}

\newcommand{\AvarUp}{A^{\rm Net}_{k}}
\newcommand{\AvarDn}{A^{\rm Int}_{k}}
\newcommand{\AvarTD}[1][k]{A\TD{#1}}
\newcommand{\AvarUpOS}{A^{\rm OS,N}_{k}}
\newcommand{\AvarDnOS}{A^{\rm OS,I}_{k}}
\newcommand{\myZq}[1]{Z\OS{#1}}

\newcommand{\Nparts}{n_p}
\newcommand{\Part}[1]{\mySet{I}_{#1}}
\newcommand{\Clust}{\mySet{J}_q^s}
\newcommand{\NegClust}{\bar{\mySet{J}}_q^s}
\newcommand{\EmpSet}{\varnothing}
\newcommand{\Nclusts}{n_{\rm cl}^q}

\long\def\symbolfootnote[#1]#2{\begingroup\def\thefootnote{\fnsymbol{footnote}}\footnote[#1]{#2}\endgroup}

\acrodef{ian}[IAN]{interference as noise}
\acrodef{td}[TD]{time-division}
\acrodef{sd}[SD]{simultanuous decoding}
\acrodef{bs}[BS]{base station}
\acrodef{mimo}[MIMO]{multiple-input multiple-output}
\acrodef{mac}[MAC]{multiple access channel}
\acrodef{tdd}[TDD]{time-division duplex}
\acrodef{ut}[UT]{user terminal}
\acrodef{cdf}[CDF]{cumulative distribution function}
\acrodef{pdf}[PDF]{probability density function}
\acrodef{ps}[PS]{pilot sequence}
\acrodef{se}[SE]{spectral efficiency}

\acrodef{dft}[DFT]{discrete Fourier transform}
\acrodef{nb}[NB]{narrowband}
\acrodef{dt}[DT]{discrete-time}
\acrodef{ct}[CT]{continuous-time}
\acrodef{evd}[EVD]{eigenvalue decomposition}
\acrodef{svd}[SVD]{singular valued decomposition}
\acrodef{soi}[SOI]{signal of interest}
\acrodef{awgn}[AWGN]{additive white Gaussian noise}
\acrodef{wss}[WSS]{wide-sense stationary}
\acrodef{mmse}[MMSE]{minimum mean-squared error}
\acrodef{mi}[MI]{mutual information}
\acrodef{lmmse}[LMMSE]{linear MMSE}
\acrodef{map}[MAP]{maximum a-posteriori probability}
\acrodef{mi}[MI]{mutual information}
\acrodef{isi}[ISI]{intersymbol interference}
\acrodef{snr}[SNR]{signal-to-noise ratio}
\acrodef{pc}[PC]{proper-complex}
\acrodef{ptp}[PtP]{point-to-point}
\acrodef{sinr}[SINR]{signal-to-interference-and-noise ratio}
\acrodef{pdf}[PDF]{probability density function}
\acrodef{taf}[TAF]{truncated amplitude flow}
\acrodef{krp}[KRP]{Khatri-Rao product}
\acrodef{rv}[RV]{random variable}
\acrodef{ic}[IC]{intereference channel}

 \IEEEoverridecommandlockouts
 
 \ifsingle
\fi %------------------
\title{On the Spectral Efficiency of Noncooperative Uplink Massive MIMO Systems
}

\vspace{-0.75cm}

\author{
	\IEEEauthorblockN{\vspace{-0.2cm} Nir Shlezinger and Yonina C. Eldar\\
	}
	\thanks{This project has received funding from the European Union’s Horizon 2020 research and innovation program under grant No. 646804-ERC-COG-BNYQ.
	The authors are with the department of EE, Technion –- Israel Institute of Technology, Haifa, Israel (nirshlezinge@technion.ac.il; yonina@ee.technion.ac.il). 	
 }
\ifsingle	
	\vspace{-2.0cm}
\else
	\vspace{-1.0cm}
\fi %-----------------
	
}

\vspace{-0.75cm}

\begin{document}

\maketitle
\pagestyle{plain}
\thispagestyle{plain}
%----------------------------------------------------------------------------------------
%	ABSTRACT
%----------------------------------------------------------------------------------------
\begin{abstract}
%	\vspace{-0.25cm}
Massive multiple-input multiple-output (MIMO) systems have been drawing considerable interest due to the growing throughput demands on wireless networks.  
In the uplink, massive MIMO systems are commonly studied assuming that each base station (BS) decodes the signals of its user terminals separately and linearly while treating all interference as noise. Although this approach provides improved spectral efficiency which scales with the number of BS antennas in favorable channel conditions, it is generally sub-optimal from an information-theoretic perspective. 
In this work we characterize the spectral efficiency of massive MIMO  when the BSs are allowed to {\em jointly} decode the received signals. In particular, we consider four  schemes for treating the interference, and derive  the achievable average ergodic rates for both finite and asymptotic number of antennas  for each scheme. 
Simulation tests of the proposed methods illustrate their gains in spectral efficiency compared to the standard approach of separate linear decoding, and show that the standard approach fails to capture the actual achievable rates of massive MIMO systems, particularly when the interference is dominant. 
%
%	{\textbf{\textit{Index terms---}} Massive MIMO, spectral efficiency.}
\end{abstract}

%----------------------------------------------------------------------------------------
%	INTRODUCTION
%----------------------------------------------------------------------------------------
\vspace{-0.4cm}
\section{Introduction}
\label{sec:Intro}
\vspace{-0.15cm}
% Paragraph - what is massive MIMO
A major challenge of future wireless systems is to meet the growing throughput demand. 
A promising method for increasing the \ac{se}  is to equip the \acp{bs} with a large number of antennas. Such systems, referred to as {\em massive \ac{mimo} systems}, were shown to provide improved throughput which is scalable with the number of \ac{bs} antennas \cite{Marzetta:15}, and are  the focus of considerable research attention in recent years. 

% Paragraph - current works on massive MIMO spectral efficiency
% Should I mention here that I focus on the uplink?
Massive \ac{mimo} systems are traditionally noncooperative multi-cell multi-user networks \cite{Lu:14}, where in each cell  a set of single-antenna \acp{ut} are served by a multi-antenna \ac{bs}. Each \ac{bs} estimates the unknown channel to its \acp{ut} in a \ac{tdd} manner prior to data transmission.
The pioneering work of Marzetta \cite{Marzetta:10} showed that, in certain favorable channel conditions and fixed number of \acp{ut} in each cell, and when the \acp{bs} perform separate linear decoding, the effects of channel estimation error and channel noise are made negligible as the number of \ac{bs} antennas increases. Furthermore, performance is limited by pilot contamination, which is the interference caused by pilot reuse among cells. 
The impact of pilot contamination on \ac{se} was further studied in \cite{Jose:11} and \cite{Fernandes:13}.
 The work \cite{Hoydis:13} characterized the \ac{se} of linear decoders under more general channel conditions, when the  number of \acp{ut} is proportional to the number of \ac{bs} antennas.  
 The tradeoff between \ac{se} and energy efficiency was studied in \cite{Ngo:13}, while  \cite{Bjornson:16} treated the effect of \ac{ut} allocation on \ac{se}. \ac{ut} allocation schemes were considered in~\cite{Bethanabhotla:16}.

% Paragraph - restrictions of previous works,
Focusing on the uplink,  namely, on the communications from the \acp{ut} to the \acp{bs}, all the works above restricted the \acp{bs} to separately decode the signal of each \ac{ut} based on some linear transformation of the channel output, such as matched filtering or \ac{mmse} filtering, while interference is treated as noise. From an information-theoretic perspective, this approach is sub-optimal, as the massive \ac{mimo} network is a set of interfering \acp{mac}. The capacity region of interfering \acp{mac} is  
unknown (In fact, even the capacity region of simple two interfering \ac{ptp} channels is generally unknown \cite[Ch. 6]{ElGamal:11}). Thus, while separate decoding and treating interference as noise is generally a sub-optimal approach for such channels \cite[Ch. 6]{ElGamal:11},  
it is not clear how far it is from optimality. 
In fact, previous studies on the gap of massive \ac{mimo} schemes from optimality assumed no intercell interference, see, e.g., \cite[Fig. 11]{Marzetta:15} and \cite[Fig. 4a]{Bjornson:16a}.  
Works studying similar channels without restricting the \acp{bs} to decode separately and treat interference as noise include 
 \cite{Girnyk:14}, which studied the achievable ergodic sum-rate of \ac{mimo} \acp{mac} with interference and  a-priori known channel in the asymptotic number of antennas regime; 
 the works  \cite{Hassibi:03,Rusek:12,Yang:13}, which studied block-fading \ac{mimo} \ac{ptp} channels; 
and  \cite{Soysal:10b}, which focused on \ac{mimo} \acp{mac} with channel estimation and without interference.

% Paragraph - in this work...
In this work we study noncooperative massive \ac{mimo} systems, focusing on the uplink, without restricting the  \acp{bs} to decode separately. In addition, we do not collectively treat interference as noise, and allow the \acp{bs} to decode the interfering signals. We characterize the \ac{se}, measured as the achievable average   ergodic rate over the entire multi-cell network,  
of three approaches for handling the intercell interference, commonly studied in the network information theoretic context of interference channels \cite[Ch. 6]{ElGamal:11}: 
%\begin{enumerate}
%	\item 
	In the first scheme, each \ac{bs} jointly decodes the signals of its corresponding \acp{ut}, and treats the intercell interference as noise.
%	\item 
	In the second scheme, each \ac{bs} decodes the signals of all the \acp{ut} in the network.
%	\item 
	In the third scheme, the data transmission phase is divided between the cells such that in each time instance only the \acp{ut} of a single cell transmit to their \ac{bs}, thus effectively canceling the intercell interference.	
%\end{enumerate} 
Note that these schemes do not treat how the \acp{ut} encode the transmitted signals, but only how the signals are decoded, and how their transmission is synchronized.  
Unlike the standard approach in the analysis of massive \ac{mimo} systems,  we allow the \acp{bs} to jointly decode the signals of their corresponding \acp{ut}. 
For each approach we first characterize the \ac{se} for a finite number of \ac{bs} antennas, and then analyze the \ac{se} in the massive \ac{mimo} regime, i.e., when the number of \ac{bs} antennas approaches infinity, using results from random matrix theory. 
Next, we study an optimized network which combines all the above schemes to maximize the  \ac{se}, by allowing  each \ac{bs} to decode some of the intercell interference while treating the rest as noise, and dividing the transmission phase such that the intercell interference is reduced but not necessarily canceled.

While these techniques are computationally more complex than the traditional approach of separate decoding and treating interference as noise, the characterization of their achievable average ergodic rate quantifies how much can be gained by removing the restrictions of the traditional approach and by properly treating massive \ac{mimo} systems as a set of interfering \acp{mac}. 
Furthermore, while the complexity of optimal joint decoding is known to grow exponentially with the number of \acp{ut}, its performance can be approached using interference cancellation \cite[Pg. 540]{Tse:05}, whose complexity only	grows linearly with the number of \acp{ut}, i.e., the same complexity order as separate linear decoding \cite{Andrews:05}, at the cost of increased decoding latency. 
Alternatively, recent developments in machine learning suggest that deep neural networks can perform accurate joint decoding at reduced complexity and latency, based on a sufficiently large training data, see, e.g., \cite{Wiesel:17}.
Consequently, the proposed analysis allows future communications engineers to understand exactly what can be gained by joint-decoding, beyond mere intuition, and accordingly to decide whether or not to implement such schemes, in light of the cost.
	%In particular, interference cancellation obtains the optimal sum-rate in \acp{mac} with Gaussian noise \cite[Pg. 540]{Tse:05}.
%	However,  interference cancellation increases the decoding latency. 
%\else
% order is the same as separate linear decoding, at the cost of increased  decoding latency \cite{Andrews:05}.
%\fi	
	  
Our numerical study demonstrates that substantial gains in \ac{se} can be obtained by allowing the \acp{bs} to perform joint decoding and by properly applying methods for handling the interference. This indicates that the approach of separately decoding a linear transformation of the channel output fails to capture the fundamental limits of massive \ac{mimo} networks. For example, we illustrate that when the intercell interference is dominant, 
a   relevant scenario for future cellular networks \cite{Ghosh:12},
the traditional approach results in a \ac{se} which approaches zero, while, when the \acp{bs} are allowed to jointly decode the interference, non-negligible average ergodic rates are achieved.

The rest of this paper is organized as follows: 
Section~\ref{sec:Preliminaries} presents the massive \ac{mimo} network model, and reviews some relevant results from random matrix theory.
Section~\ref{sec:Rates}  derives the \ac{se} of the considered schemes. % and details the combined optimized approach. 
Section~\ref{sec:Simulations} provides simulation examples. 
Finally,  Section~\ref{sec:Conclusions}  concludes the paper. 
Proofs of the  results stated in the paper 
are detailed in the appendix.

Throughout the paper, we use   
boldface lower-case letters for vectors, e.g., ${\myDetVec{x}}$; 
the $i$-th element of ${\myDetVec{x}}$ is written as $(\myDetVec{x})_i$.  
Matrices are denoted with boldface upper-case letters,  e.g.,  
$\myDetMat{m}$, and we use $(\myDetMat{m})_{i,j}$ to denote its $(i,j)$-th element. We use $\myI_{n}$ to denote the $n \times n$ identity matrix.   
Hermitian transpose, transpose, complex conjugate, stochastic expectation,  and mutual information are written as $(\cdot)^H$, $(\cdot)^T$, $(\cdot)^*$,  $\E\{ \cdot \}$,  and $I\left( \cdot ~ ; \cdot \right)$, respectively. 
$\delta_{k,l}$ is the Kronecker delta, i.e., $\delta_{k,l}\! =\! 1$ when $k\!=\!l$ and $\delta_{k,l}\! =\! 0$ otherwise. 
%
%Let $\myMat{0}_{a\times b}$ denote the all-zero $a \times b$ matrix.
%
We use ${\rm {Tr}}\left(\cdot\right)$ to denote the trace operator, 
$\otimes$ is the Kronecker product,  
$\Dist$ denotes equality in distribution of two \acp{rv}, and $\mySet{C}$ is the set of complex numbers.  
Unless stated otherwise, all logarithms are taken to base-2. 
Finally, for an $n \! \times \! n$ matrix $\myDetMat{x}$,  $\myDetVec{x}\! =\! {\rm vec}\left(\myDetMat{x}\right)$ is the $n^2 \! \times \! 1$ column vector obtained by stacking the columns of $\myDetMat{x}$ one below the other.  
The matrix $\myDetMat{x}$ is recovered from $\myDetVec{x}$ via $\myDetMat{x} = {\rm vec}^{-1} \!\left(\myDetVec{x}\right)$. 

%Lastly, for any sequence, possibly multivariate, $\myVec{q}[i]$, $i \in \mySet{Z}$, and integers $a_1 < a_2$, we use $\myVec{q}_{a_1}^{a_2}$ to denote the column vector obtained by stacking $\left[\myVec{q}[a_1]^T,\ldots,\myVec{q}[a_2]^T\right]^T$ and $\myVec{q}^{a_2} \equiv \myVec{q}_{0}^{a_2}$.

%----------------------------------------------------------------------------------------
%	PRELIMINARIES
%----------------------------------------------------------------------------------------
\vspace{-0.2cm}
\section{Preliminaries and System Model}
\label{sec:Preliminaries} 
%-----------------------------------
%	Problem Formulation
%----------------------------------- 
\subsection{Problem Formulation}
\label{subsec:Pre_Problem} 
We consider a noncooperative multi-cell multi-user \ac{mimo} system with $\Ncells$ cells, focusing on the uplink. 
In each cell, a \ac{bs} equipped with $\Nantennas$ antennas serves $\Nusers$ single-antenna \acp{ut}. 
We assume that  $\Nantennas$ and $\Nusers$ are sufficiently large to carry out large scale (asymptotic) analysis, and fix the ratio of  the number of \acp{ut} to the number of antennas $\AntRatio \triangleq \frac{\Nusers}{\Nantennas}$. 

% TODO NIR - cite references for channel model
Let $\Dmat_{k,l}$ be an $\Nusers \times \Nusers$ random diagonal matrix with positive diagonal entries $\{\dcoeff_{k,l,m} \}_{m=1}^{\Nusers}$ representing the attenuation between the $m$-th \ac{ut} of the $l$-th cell and the $k$-th \ac{bs}, $k,l \in \{1,2,\ldots,\Ncells\} \triangleq\NcellsSet$. We assume that the attenuation coefficients are mutually independent, and that for a fixed $k,l$, the attenuation coefficients from the \acp{ut} of the $l$-th cell and the $k$-th \ac{bs}, $\{\dcoeff_{k,l,m} \}_{m=1}^{\Nusers}$, are also identically distributed. 
Furthermore, let $\Hmat_{k,l}\in\mySet{C}^{\Nantennas\times\Nusers}$ be a random proper-complex\footnote{{\label{ftn:Proper}Following \cite[Def. 1]{Nesser:93}, we use the term {\em proper-complex} for complex-valued random vectors and matrices whose pseudo-covariance vanishes, thus their second-order statistical moment is completely characterized by the covariance matrix.}} zero-mean Gaussian matrix with i.i.d. entires of unit variance, representing the instantaneous channel response between the \acp{ut} of the $l$-th cell and the $k$-th \ac{bs},  $k,l \in \NcellsSet$. 
 For each $(k_1,l_1) \ne (k_2,l_2)$, $\Hmat_{k_1,l_1}$ and $\Hmat_{k_2,l_2}$ are mutually independent, and are also independent of  $\{\Dmat_{k,l} \}_{k,l \in \NcellsSet}$.
Let $\Gmat_{k,l} = \Hmat_{k,l} \Dmat_{k,l}$ be the random channel matrix from the \acp{ut} in the $k$-th cell to the $l$-th \ac{bs}. 
We assume a block-fading model for $\{\Hmat_{k,l} \}_{k,l \in \NcellsSet}$, in which the channel coefficients $\{\Hmat_{k,l} \}_{k,l \in \NcellsSet}$ are unknown and remain constant only for a coherence duration of $\Tcoh$ symbols. As in, e.g.,  \cite{Bjornson:16}, each \ac{bs} knows its  
corresponding  attenuation coefficients\footnote{Although the attenuation coefficients are assumed to vary slowly, we do not assume that they are {\em slow-fading}, as we allow the codewords to span a sufficiently large number of independent realizations of $\{\Dmat_{k,l} \}_{k,l \in \NcellsSet}$.}  
i.e., the $k$-th \ac{bs} knows $\{\Dmat_{k,l} \}_{l \in \NcellsSet}$.
Let $\myW_k[i]\in \mySet{C}^{\Nantennas}$, $k \in \NcellsSet$, be an i.i.d. zero-mean proper-complex Gaussian signal with  covariance matrix $\SigW \myI_{\Nantennas}$, $\SigW > 0$, representing the additive channel noise at the $k$-th \ac{bs}.  

Channel estimation is carried out in a \ac{tdd} fashion, where the coherence duration $\Tcoh$ is divided into a channel estimation phase, consisting of $\Tpilots$ pilot symbols,  
and a data transmission phase, consisting of $\Tdata = \Tcoh  - \Tpilots$ data symbols. 
During the channel estimation phase, each \ac{ut} sends a deterministic orthogonal \ac{ps}, where the \acp{ps} are the same in all cells. 
The \acp{bs} use the a-priori knowledge of the \acp{ps} to estimate the channel.
Letting $s_m[i]$ denote the $i$-th pilot symbol of the $m$-th user in each cell, $m \in \{1,2,\ldots,\Nusers\} \triangleq \NusersSet$, $i \in \{1,2,\ldots, \Tpilots\}$, 
and defining $\myS[i] \triangleq [s_1[i], s_2[i], \ldots, s_{\Nusers}[i]]^T$, 
the channel output at the $k$-th \ac{bs}, $k \in \NcellsSet$, is given by 
\begin{equation}
\label{eqn:Channel_Est1}
\myY_k[i] = \sum\limits_{l=1}^{\Ncells}\Gmat_{k,l}\myS[i] + \myW_k[i], \qquad i = 1,2,\ldots, \Tpilots. 
\end{equation}  
The orthogonality of the \acp{ps} implies that for all $m_1,m_2 \in \NusersSet$,
$\sum\limits_{i=1}^{\Tpilots}s_{m_1}[i]s_{m_2}^*[i] = \Tpilots \cdot \delta_{m_1,m_2}$. Furthermore, the \ac{ps} length, $\Tpilots$, must not be smaller than the number of \acp{ut}, $\Nusers$ \cite[Sec. III-A]{Marzetta:10}. 

During data transmission, we assume equal unit power Gaussian codebooks among all \acp{ut}, i.e., the transmitted signal of the \acp{ut} in the $k$-th cell, $k \in \NcellsSet$, denoted $\myX_k[i]\in\mySet{C}^{\Nusers}$, is a zero-mean Gaussian vector with identity covariance.
%consists of i.i.d. zero-mean unit variance proper-complex Gaussian \acp{rv}.
%is a Gaussian random vector with i.i.d. zero-mean unit variance entries. 
The channel output at the $k$-th \ac{bs} is given by 
\begin{equation}
\label{eqn:Channel_Data1}
\myY_k[i] = \sum\limits_{l=1}^{\Ncells}\Gmat_{k,l}\myX_l[i] + \myW_k[i], \qquad i \!=\! \Tpilots\!+\!1,\Tpilots\!+\!2,\ldots, \Tcoh, 
\end{equation}
where  $\{\myX_l[i]\}_{l \ne k}$ %$\sum\limits_{l=1, l \ne k}^{\Ncells}\Gmat_{k,l}\myX_l[i]$ 
represents the intercell interference. 

\label{txt:QuasiStatic}
Our goal is to characterize the \ac{se} of noncooperative multi-cell multi-user \ac{mimo} systems, represented as the achievable average ergodic rate. Letting $\R{k,m}$ be the achievable ergodic rate of the $m$-th \ac{ut} in the $k$-th cell, the  achievable average ergodic rate is defined as 
\begin{equation}
\label{eqn:ErgSumRate}
\R{} 
\triangleq  \frac{\Tdata}{\Tcoh}\cdot\frac{1}{\Ncells \cdot \Nusers}\sum\limits_{k=1}^{\Ncells}\sum\limits_{m=1}^{\Nusers}  \R{k,m},
\vspace{-0.1cm}
\end{equation} 
where the factor $\frac{\Tdata}{\Tcoh} = 1 - \frac{\Tpilots}{\Tcoh}$ follows since only $\Tdata$ symbols of each coherence interval are used for data transmission. 
 	Each $\R{k,m}$ is computed by averaging the achievable rate over a large number of independent realizations of the attenuation coefficients $\{\Dmat_{k,l} \}_{k,l \in \NcellsSet}$.	This approach corresponds to quasi-static capacity analysis, which assumes multiple long transmission bursts, where the \ac{se} is computed assuming that the  attenuation coefficients do not change during each burst, see \cite[Sec. 4]{Foschini:98}. The resulting \ac{se} characterization yields a tight upper bound to the throughput of a practical code with codelength that is smaller than the  coherence time of the  attenuation coefficients.
In particular, we study the \ac{se} in the {\em massive \ac{mimo} regime}, namely, when the number of \ac{bs} antennas, $\Nantennas$, grows infinitely large while $\AntRatio$, which denotes the ratio of the number of \acp{ut}, $\Nusers$, to the number of \ac{bs} antennas, is kept fixed and finite. As explained in \cite[Sec. 3]{Hoydis:13}, this asymptotic analysis provides tight approximations of the \ac{se} of practical massive \ac{mimo} systems, where both $\Nantennas$ and $\Nusers$ are large yet finite. This setup is different from that considered in \cite{Marzetta:10}, where only $\Nantennas$ is assumed to be arbitrarily large.  

The standard approach in the massive \ac{mimo} literature, e.g., \cite{Marzetta:10,Fernandes:13,Ngo:13,Jose:11,Hoydis:13, Bjornson:16}, is to  restrict the \acp{bs} to separately decode the signal of each \ac{ut} from some linear transformation of the channel output. We henceforth refer to this approach as {\em separate linear decoding}. Here, in order to recover the symbol of the $m$-th \ac{ut} in the $k$-th cell, the \ac{bs} computes the inner product between the received vector $\myY_k[i]$ and some linear filter $\myDetVec{q}_{k,m} \in \mySet{C}^{\Nantennas}$, and uses the result to decode only the symbol of the $m$-th \ac{ut}. 
Letting $\gamma_{k,m}$ be an \ac{rv} representing the \ac{sinr} of the channel relating the $m$-th \ac{ut} of the $k$-th cell and its corresponding \ac{bs}, $k \in \NcellsSet, m \in \NusersSet$, the \ac{se} of this approach is given by 
\begin{equation}
\label{eqn:SepLinRate}
\RSEP{}
=  \frac{\Tdata}{\Tcoh}\cdot\frac{1}{\Ncells \cdot \Nusers}\sum\limits_{k=1}^{\Ncells}\sum\limits_{m=1}^{\Nusers}  
\E \left\{\log \left( 1 + \gamma_{k,m}\right)  \right\}. 
\end{equation} 
The stochastic expectation in \eqref{eqn:SepLinRate} is carried out with respect to the \ac{sinr} \ac{rv} $\gamma_{k,m}$.  The \ac{sinr} is determined by the filter $\myDetVec{q}_{k,m}$, the attenuation coefficients $\{\Dmat_{k,l} \}_{l \in \NcellsSet}$, and the noise power $\SigW$, see, e.g.,  \cite[Sec. II]{Hoydis:13}. The randomness of the \ac{sinr} follows since the filter $\myDetVec{q}_{k,m}$  depends on the (random) estimated channel, and from the randomness of the attenuation coefficients $\{\Dmat_{k,l} \}_{l \in \NcellsSet}$. 
The novel aspect of our analysis is that {\em we allow the \acp{bs} to use joint multi-user detection}. 
While multi-user detection is inherently more complex than separate linear decoding, especially for a large number of \acp{ut}, the resulting analysis captures the fundamental properties of noncooperative massive \ac{mimo} systems, and quantifies how much is lost, in terms of \ac{se}, due to the restriction to use separate linear decoding. 
Furthermore, we emphasize that the additional complexity is required only at the \acp{bs}, i.e., no additional processing is required at the \acp{ut}. Finally, the performance of optimal multi-user detection can be approached at a significantly reduced complexity using deep learning algorithms, as indicated in \cite{Wiesel:17}. Alternatively, optimal multi-user detection can be implemented using iterative algorithms, whose complexity only grows linearly with the number of \acp{ut}, at the cost of increased decoding delay, see, e.g., \cite{Andrews:05}.

%-----------------------------------
%	The Marchenko-Pastur Law for Gaussian Diagonal Random Matrices
%-----------------------------------
\vspace{-0.2cm}
\subsection{Results from Large Random Matrix Theory}
\label{subsec:Pre_Marcenko}
\vspace{-0.1cm}
In our study  we rely on some existing results from the theory of large random matrices. 
To formulate these results, we first recall the definition of the empirical eigenvalue \ac{cdf}: 
For an $\Nantennas \times \Nantennas$ random Hermitian matrix $\myRandMat{A}$ with eigenvalues $\{\lambda_i\left( \myRandMat{A}\right)\}_{i=1}^{\Nantennas}$, the (random) empirical \ac{cdf} of its eigenvalues is given by 
$\CDF{\myRandMat{A}}(x) = \frac{1}{\Nantennas}\sum\limits_{i=1}^{\Nantennas} 1\left\{\lambda_i\left( \myRandMat{A}\right) \le x  \right\}$, 
where  $1\{\cdot\}$ is the indicator function. Note that $\CDF{\myRandMat{A}}(x)$ is a random function of the real scalar $x$.
The following result, which is obtained from the Mar\v{c}enko-Pastur law for the asymptotic eigenvalue distribution of large random matrices \cite{Marchenko:67}, is frequently used in our analysis:
\begin{theorem}
	\label{thm:MarPast1}
	{\bf \cite[Thm. 2.39]{Tulino:04}:} 
	Let $\Hmat \in \mySet{C}^{\Nantennas \times \Nusers}$ be a proper-complex random matrix with i.i.d. entries with zero-mean and unit variance, and let $\myRandMat{A} \in \mySet{C}^{\Nusers \times \Nusers}$ be a Hermitian non-negative random matrix, independent of $\Hmat$, whose empirical	eigenvalue \ac{cdf} converges almost surely to the nonrandom \ac{cdf} of the 
	real-valued non-negative 
	scalar \ac{rv} $A$. 
	Then, for  fixed $\frac{\Nusers}{\Nantennas} = \AntRatio$, we have that as $\Nantennas \rightarrow \infty$, 
	\vspace{-0.1cm}
	\begin{align}
	&\frac{1}{\Nantennas}\log \left|\myI_{\Nantennas} + \frac{1}{\Nantennas} \Hmat \myRandMat{A} \Hmat^H \right| \notag \\
	&\qquad \AsConv
	 \AntRatio \cdot \E\left\{\log \left(1 + \myEta \cdot A \right)  \right\} - \log \myEta + (\myEta - 1) \log e
	 \notag \\
	& \qquad
	\triangleq \MPFunc(A, \AntRatio), 
	\label{eqn:MarPast1}
	\vspace{-0.1cm}
	\end{align}
	where $\AsConv$ denotes almost sure convergence, and  $\myEta \in (0,1]$ is the solution to 
%	\vspace{-0.1cm}
%	\begin{equation}
	$\AntRatio = \frac{1- \myEta}{1 - \E\left\{ \frac{1}{1 + \myEta \cdot A}  \right\}}$.
%	\label{eqn:EtaDef}
%	\vspace{-0.1cm}
%	\end{equation}
\end{theorem}

We note that when $\myRandMat{A}$ is the deterministic matrix $\myI_{\Nusers}$, \eqref{eqn:MarPast1} specializes to the limit in \cite[Eq. (1.14)]{Tulino:04}, which characterizes the asymptotic capacity of Rayleigh fading \ac{ptp} \ac{mimo} channels.  
Furthermore, as the left-hand side of \eqref{eqn:MarPast1} is a non-negative real-valued \ac{rv}, the deterministic function $ \MPFunc(A, \AntRatio) $ is also non-negative real-valued.   

%----------------------------------------------------------------------------------------
%	Average Achievable Rates
%---------------------------------------------------------------------------------------- 
\section{Achievable Average Ergodic Rates}
\label{sec:Rates} 
In order to compute the \acp{se}, namely, the achievable average  ergodic rates, we recall 
 that the uplink massive \ac{mimo} system is inherently a set of interfering \acp{mac}. 
In particular, in \eqref{eqn:Channel_Data1},  $\myY_k[i]$ is the \ac{mac} output, the entries of $\myX_k[i]$ are the \ac{mac} inputs, and  $\{\myX_l[i]\}_{l \ne k}$ is the interference.  
 Consequently, we consider the following common approaches for handling the intercell interference: treating intercell interference as noise, simultaneous decoding, and time division between cells. 
 The first two schemes determine only how each \ac{bs} treats the intercell interference when decoding its input, while the third approach eliminates the intercell interference without modifying the transmitted signals, by synchronizing the cells to avoid simultaneous transmission. We emphasize that these methods do not treat how the transmitted data is encoded. 
 
 To study these approaches, we first elaborate on the channel estimation phase in Subsection \ref{subsec:ChEst}. Then, in Subsections \ref{subsec:IAN}--\ref{subsec:TD}, we discuss each method and its \ac{se} for a finite number of \ac{bs} antennas and in the massive \ac{mimo} regime. 
  Unlike previous works, e.g., \cite{Marzetta:10,Hoydis:13,Fernandes:13,Jose:11,Ngo:13,Bjornson:16}, we do not restrict our attention to separate linear decoding, and allow the \acp{bs} to jointly decode the signals of their \acp{ut}.
 The proofs of our results follow the same outline for each approach:
 \begin{itemize}
 	\item To characterize the \ac{se} for a finite number of \ac{bs} antennas we first divide the received signal into a signal which the \ac{bs} decodes and an uncorrelated signal which is considered as noise. Then, we compute the correlation matrix of the equivalent noise, and use worst-case uncorrelated noise arguments, see, e.g., \cite{Hassibi:03}, to obtain an expression for the \ac{se}.
 	\item To characterize the \ac{se} in the massive \ac{mimo} regime, we prove that  the expression for the \ac{se} for a finite number of \ac{bs} antennas satisfies the conditions of Theorem \ref{thm:MarPast1}. Then, we apply Theorem \ref{thm:MarPast1} to explicitly obtain the \ac{se} in the massive \ac{mimo} regime.
 \end{itemize}
The detailed proofs are relegated to the appendix.  
 Next, in Subsection \ref{subsec:Example}, we provide an illustrative example for which we analytically compare the \acp{se} of the considered approaches. In particular, this example indicates that treating interference as noise is the best approach when the intercell interference is weak, while simultaneous decoding is the best approach when the interference is dominant.
 Finally, in Subsection \ref{subsec:Optimize}, we propose a method for combining the schemes for handling the intercell interference such that  the  \ac{se} is optimized.

% TODO NIR FINISH VERIFY FROM HERE
%-----------------------------------
%	Channel Estimation
%----------------------------------- 
\subsection{Channel Estimation}
\label{subsec:ChEst} 
As stated in the system model, the first $\Tpilots$ symbols of each coherence interval are orthogonal \acp{ps}  used by the \acp{bs} to produce the \ac{mmse} estimate of their corresponding channel responses.  
Define the $\Nantennas \times \Tpilots$ random matrices $\myYmat_k \triangleq \big[\myY_k[1],\ldots,\myY_k[\Tpilots] \big]$, $\myWmat_k \triangleq \big[\myW_k[1],\ldots,\myW_k[\Tpilots] \big]$, and the $\Nusers \times \Tpilots$ deterministic matrix  $\mySmat \triangleq \big[\myS[1],\ldots,\myS[\Tpilots] \big]$. 
From \eqref{eqn:Channel_Est1} we have that for all $k \in \NcellsSet$:  
\begin{equation}
\label{eqn:Channel_Training_MarForm}
 \myYmat_k = \sum\limits_{l=1}^{\Ncells}\Gmat_{k,l} \mySmat + \myWmat_k. 
\end{equation}
Since the \acp{ps} are orthogonal and $\Tpilots \ge \Nusers$, we have that $\mySmat \mySmat^H = \Tpilots \cdot \myI_{\Nusers}$. 
Let $\UHmat$ be an $\Nantennas \times \Nusers$ zero-mean proper-complex Gaussian random matrix with i.i.d. unit variance entries independent of  $\{\Dmat_{k,l} \}_{k,l \in \NcellsSet}$,  and define the \acp{rv}  
\begin{equation}
\bcoeff_{k,l,m} \triangleq \frac{\Tpilots \dcoeff_{k,l,m}^2}{\SigW + \Tpilots\sum\limits_{l'=1}^{\Ncells} \dcoeff_{k,l',m}^2 }, \quad k,l \in \NcellsSet, m \in \NusersSet,
\label{eqn:BmatDef} 
\end{equation}
and the $\Nusers \times \Nusers$ diagonal matrices $\{\Bmat_{k,l}\}_{k,l \in \NcellsSet}$ with diagonal entries 
 $\{\bcoeff_{k,l,m} \}_{m=1}^{\Nusers}$. The \ac{mmse} channel estimate  and its statistical characterization are stated in the following lemma:
\begin{lemma}
	\label{lem:ChEstLem}
	The \ac{mmse} estimate of $\Gmat_{k,l}$ from $ \myYmat_k $ and $\{\Dmat_{k,l}\}_{l \in \NcellsSet}$ is given by 
		\begin{equation}
		\label{eqn:ChEstLem1}
		\hat{\Gmat}_{k,l} = \Tpilots^{-1}\myYmat_k \mySmat^H  \Bmat_{k,l}. 
		\end{equation}
		Furthermore,  the \ac{mmse} estimate $\hat{\Gmat}_{k,l}$  is distributed as $\hat{\Gmat}_{k,l} \Dist \UHmat\Bmat_{k,l}^{1/2}\Dmat_{k,l}$ and its estimation error $\tilde{\Gmat}_{k,l} \triangleq {\Gmat}_{k,l} - \hat{\Gmat}_{k,l}$ is distributed as  $\tilde{\Gmat}_{k,l} \Dist \UHmat\left( \myI_{\Nusers} - \Bmat_{k,l}\right)^{1/2}\Dmat_{k,l}$.
\end{lemma}
 
	{\em Proof:}
	See Appendix \ref{app:Proof1}. 

\smallskip
The remaining $\Tdata = \Tcoh - \Tpilots$ symbols of each coherence interval are used for uplink data transmission. 
In the following subsections we study the achievable average ergodic rates of several schemes using the \ac{mmse} channel estimates  \eqref{eqn:ChEstLem1}. 

%-----------------------------------
%	Interference as Noise
%----------------------------------- 
\subsection{Decoding Scheme 1 - Interference as Noise}
\label{subsec:IAN} 
We first study  the \ac{se} when each \ac{bs} treats the intercell interference as noise. 
\textcolor{NewColor}{The intuition here is that the \acp{bs} only decode their relevant messages, thus the transmission rate of each \ac{ut} should only guarantee reliable decoding by its corresponding \ac{bs}.}
 In particular, the $k$-th \ac{bs}, $k \in \NcellsSet$, jointly decodes the signals transmitted by the \acp{ut} associated with the $k$-th cell, $\myX_k[i]$, and treats  the signals transmitted by all \acp{ut} which are not associated with  the $k$-th cell, $\{\myX_l[i]\}_{l \ne k}$, %, $\{\myX_j[i]\}_{j \ne k}$, 
as noise.   
The fundamental difference between the decoding scheme considered here and previous works on massive \ac{mimo} systems, e.g., \cite{Marzetta:10,Fernandes:13,Ngo:13,Jose:11,Hoydis:13, Bjornson:16}, which also assumed that the \acp{bs} treat intercell interference as noise, is that these works restricted each \ac{bs} to decode the signals transmitted from each of its associated \acp{ut} {\em separately}, thus the channel from the \acp{ut} to the \ac{bs} is treated as a set of \ac{ptp} channels, and the focus is on characterizing the  \ac{sinr} of the channel from each \ac{ut} to its \ac{bs}. Here, we allow the \acp{bs} to {\em jointly} decode  the signals transmitted by their \acp{ut},  exploiting the fact that the channel from the \acp{ut} to their associated \ac{bs} is a  \ac{mac}.

Using the \ac{mmse} channel estimate $\hat{\Gmat}_{k,k}$ and its estimation error $\tilde{\Gmat}_{k,k}$, the received signal  at the $k$-th \ac{bs} during data transmission  
\eqref{eqn:Channel_Data1} can be written as 
\begin{equation}
\label{eqn:IANModel1}
\myY_k[i] \!=\! \hat{\Gmat}_{k,k}\myX_k[i] + \tilde{\Gmat}_{k,k}\myX_k[i] + \!\!\sum\limits_{l=1, l\ne k}^{\Ncells}\!\!\!\Gmat_{k,l}\myX_l[i] \!+\! \myW_k[i]. 
\end{equation}
By treating interference as noise, the equivalent noise signal is defined as 
$\myV\IAN{k}[i] \triangleq \tilde{\Gmat}_{k,k}\myX_k[i] + \sum\limits_{l=1, l\ne k}^{\Ncells}\Gmat_{k,l}\myX_l[i] + \myW_k[i]$, and the received signal can be written as 
\begin{equation}
\label{eqn:IANModel2}
\myY_k[i] = \hat{\Gmat}_{k,k}\myX_k[i] + \myV\IAN{k}[i], \quad i \!=\! \Tpilots\!+\!1,\Tpilots\!+\!2,\ldots, \Tcoh. 
\end{equation}
To formulate the  achievable average ergodic rate of \eqref{eqn:IANModel2}, define the \ac{rv}  
\begin{equation}
\label{eqn:TkDef}
\myT_k \triangleq  \sum\limits_{l=1 }^{\Ncells} {\rm Tr}\big(( \myI_{\Nusers} - \Bmat_{k,l} ) \Dmat_{k,l}^2 \big)  +\SigW,  
\end{equation}
and the $\Nusers \times \Nusers$ random diagonal matrices  
\begin{subequations}
	\label{eqn:AggMatDefIAN}
\begin{equation}
\AmatUp \triangleq \myT_k^{-1} \Bmat_{k,k}\Dmat_{k,k}^{-2}\sum\limits_{l=1}^{\Ncells}\Dmat_{k,l}^{4}; 
\end{equation}
and
\begin{equation}
\label{eqn:AggMatDefIAN2}
\AmatDn \triangleq \myT_k^{-1} \Bmat_{k,k}\Dmat_{k,k}^{-2}\sum\limits_{l=1, l \ne k}^{\Ncells}\Dmat_{k,l}^{4}.
\vspace{-0.2cm}
\end{equation}
\end{subequations}
The \ac{se} in the finite number of antennas regime is stated in the following proposition:
\begin{proposition}
	\label{lem:IAN1}
	 When the \acp{bs} treat intercell interference as noise, the following average ergodic rate is achievable: 
	 \begin{align}
	 \RIAN{\Nantennas}  
	 = \frac{\Tdata}{\Tcoh}\cdot\frac{1}{\Ncells \cdot \Nusers}&\sum\limits_{k=1}^{\Ncells} 
	 \Big(\E \left\{\log \left|\myI_{\Nantennas} + \UHmat \AmatUp\UHmat^H  \right|  \right\} \notag \\ 
	 &- \E \left\{\log \left|\myI_{\Nantennas} + \UHmat \AmatDn\UHmat^H  \right|  \right\} \Big), 
	 \label{eqn:IAN1}
	 \end{align}
	 where the expectations are carried out with respect to the random matrices $\UHmat$ and $\{\AmatUp, \AmatDn\}_{k \in \NcellsSet}$. 
\end{proposition}

{\em Proof:}
See Appendix \ref{app:Proof2}.

\smallskip
Next, we use Proposition \ref{lem:IAN1} to characterize the  achievable average  ergodic rate  in the massive \ac{mimo}  
regime. To that aim, define the following \acp{rv}  
\begin{subequations}
\label{eqn:AIANkDef1}
\begin{equation} 
\AvarUp \!\triangleq\! 
%\left( {\AntRatio  \sum\limits_{l=1}^{\Ncells} \E\{\left(1 - \bcoeff_{k,l,1} \right)\dcoeff_{k,l,1}^2 \} }\right)^{-1}\!\!\!\! {\bcoeff_{k,k,1}\dcoeff_{k,k,1}^2}.
%
\frac{\bcoeff_{k,k,1}\dcoeff_{k,k,1}^{-2} \sum\limits_{l=1}^{\Ncells}\dcoeff_{k,l,1}^{4}} {\AntRatio  \sum\limits_{l=1}^{\Ncells} \E\{\left(1 - \bcoeff_{k,l,1} \right)\dcoeff_{k,l,1}^2 \} }; 
\end{equation}
and 
\begin{equation}
\AvarDn\!\triangleq\! 
%\left( {\AntRatio  \sum\limits_{l=1}^{\Ncells} \E\{\left(1 - \bcoeff_{k,l,1} \right)\dcoeff_{k,l,1}^2 \} }\right)^{-1}\!\!\!\! {\bcoeff_{k,k,1}\dcoeff_{k,k,1}^2}.
%
\frac{\bcoeff_{k,k,1}\dcoeff_{k,k,1}^{-2} \sum\limits_{l=1, l\ne k}^{\Ncells}\dcoeff_{k,l,1}^{4}} {\AntRatio  \sum\limits_{l=1}^{\Ncells} \E\{\left(1 - \bcoeff_{k,l,1} \right)\dcoeff_{k,l,1}^2 \} }, 
%\label{eqn:AIANkDef2}
\end{equation}
\end{subequations}
for $k \in \NcellsSet$. 
%and let $\myEta\IAN{k} \in (0,1]$ be the solutions to 
%\begin{equation}
%\AntRatio = \frac{1- \myEta\IAN{k}}{1 - \E\left\{ \frac{1}{1 + \myEta\IAN{k} \cdot A\IAN{k}}  \right\}}, \qquad k \in \NcellsSet.
%\label{eqn:EmpAsym3}
%\end{equation} 
%
Letting  $\Nantennas \rightarrow \infty$ in  \eqref{eqn:IAN1} while fixing $\frac{\Nusers}{\Nantennas} = \AntRatio$, we obtain the  achievable average ergodic rate in the massive \ac{mimo} regime, stated in the following theorem:
\begin{theorem}
	\label{thm:IAN2}
	In the massive \ac{mimo} regime, the following average ergodic rate is achievable when treating intercell interference as noise:  
\begin{align}
\RIAN{} &\triangleq \mathop{\lim}\limits_{\stackrel{\Nantennas \rightarrow \infty}{\frac{\Nusers}{\Nantennas} = \AntRatio}}\RIAN{\Nantennas} \notag \\ 
&= \frac{\Tdata}{\Tcoh}\cdot\frac{1}{\Ncells  \cdot \AntRatio} \sum\limits_{k=1}^{\Ncells} \MPFunc\left(\AvarUp, \AntRatio \right) -  \MPFunc\left(\AvarDn, \AntRatio \right),
\label{eqn:IAN2} 
\end{align}
where  $\MPFunc\left( \cdot, \cdot\right)$ is defined in \eqref{eqn:MarPast1}.
\end{theorem}

{\em Proof:}
See Appendix \ref{app:Proof3}.

 \smallskip 
As detailed in Appendix \ref{app:Proof2}, Proposition \ref{lem:IAN1} is proved by computing the maximal achievable average ergodic rate, assuming that the equivalent noise $\myV\IAN{k}$ is Gaussian. In the standard approach of separate linear decoding, this equivalent noise is also assumed to be Gaussian, and the \ac{se}, given in  \eqref{eqn:SepLinRate}, is computed assuming that the decoder filters the received signal in \eqref{eqn:IANModel1} and decodes each entry separately. Consequently, the \ac{se} of the standard approach is always upper bounded by the \ac{se} in \eqref{eqn:IAN1} and \eqref{eqn:IAN2}.   
In the example presented in Subsection \ref{subsec:Example} and in the numerical study detailed in Section \ref{sec:Simulations} we demonstrate that the approach of treating intercell interference as noise is most beneficial when the intercell interference is weak, in agreement with the theory of two-user Gaussian interference channels \cite[Ch. 6.4.3]{ElGamal:11}.

%-----------------------------------
%	Simultanuous Decoding
%----------------------------------- 
\subsection{Decoding Scheme 2 - Simultaneous Decoding}
\label{subsec:SD} 
The opposite approach to treating interference as noise is  to decode the intercell interference.  
Specifically, each \ac{bs} now jointly decodes the signals transmitted by all \acp{ut} in the  network.  
	The rationale of this scheme is that, by decoding the intercell interference, each \ac{bs} can cancel its effect when decoding the desired messages of its corresponding \acp{ut}. However, it requires each \ac{ut} to set its rate such that its message can be reliably decoded by all the \acp{bs} in the network. 
This approach is known to be optimal in the two-user Gaussian interference channel with strong interference \cite[Ch. 6.4.2]{ElGamal:11}, and thus we expect it to achieve the best performance in networks where many \acp{ut} are not allocated to the \acp{bs} with best connectivity (a scenario which is not uncommon in wireless networks \cite{Bethanabhotla:16}).
Consequently, while this approach is more computationally complex than treating interference as noise, deriving its \ac{se} gives an indication of the fundamental performance limits of wireless networks with strong intercell interference, which cannot be obtained using the standard approach of treating interference as noise. 

From  \eqref{eqn:BmatDef} and \eqref{eqn:ChEstLem1}, it follows that $\hat{\Gmat}_{k,l} = \hat{\Gmat}_{k,k} \Dmat_{k,k}^{-2}\Dmat_{k,l}^2$. Thus, given  $\{\Dmat_{k,l}\}_{l \in \NcellsSet}$, %for the $k$-th \ac{bs}, 
obtaining the \ac{mmse} estimate of all cross-cell channels, $\{\hat{\Gmat}_{k,l}\}_{l\in\NcellsSet} $, is equivalent to obtaining only $\hat{\Gmat}_{k,k}$, and no additional pilots are required. The received signal at the $k$-th \ac{bs} %during data transmission 
\eqref{eqn:Channel_Data1} can be written as 
\begin{equation}
\label{eqn:SDModel1}
\myY_k[i] =  \sum\limits_{l=1}^{\Ncells}\hat{\Gmat}_{k,l}\myX_l[i] + \sum\limits_{l=1}^{\Ncells}\tilde{\Gmat}_{k,l}\myX_l[i]+ \myW_k[i]. 
\end{equation} 
When decoding the intercell interference along with the data, the equivalent noise is 
$\myV\SD{k}[i] \triangleq \sum\limits_{l=1}^{\Ncells}\tilde{\Gmat}_{k,l}\myX_l[i]+ \myW_k[i]$, and the received signal can be written as 
\begin{equation}
\label{eqn:SDModel2}
\myY_k[i] = \hat{\Gmat}_{k,k}{\Dmat}_{k,k}^{-2}\sum\limits_{l=1}^{\Ncells}{\Dmat}_{k,l}^2\myX_l[i] + \myV\SD{k}[i],  
\end{equation}
$i = \Tpilots+1,\Tpilots+2,\ldots, \Tcoh.$
The \ac{se} for finite $\Nantennas$ of the proposed approach is stated in the following proposition:
\begin{proposition}
	\label{lem:SD1}
	When each \ac{bs} decodes the intercell  
	interference along with the data signal,  
	 the following average ergodic rate is achievable: 
	\begin{equation}
	\label{eqn:SD1}
	\RSD{\Nantennas}  
	\!=\! \frac{\Tdata}{\Tcoh}\cdot\frac{1}{\Ncells \cdot \Nusers}\mathop{\min}\limits_{k \in \NcellsSet}
	 \left( 
	\E \left\{\log \left|\myI_{\Nantennas}\! +\! \UHmat \AmatUp\UHmat^H  \right|  \right\}\right),  
	\end{equation}
	where %$\AmatUp$ is defined in \eqref{eqn:AggMatDefIAN1}, and 
	the expectations are carried out with respect to the random matrices $\UHmat$ and $\{\AmatUp\}_{k \in \NcellsSet}$. 
\end{proposition}

	{\em Proof:}
	See Appendix \ref{app:Proof4}. 

\smallskip
Next, we use Proposition \ref{lem:SD1} to characterize the  achievable average ergodic rate in the massive \ac{mimo} regime.  
Letting  $\Nantennas \rightarrow \infty$ in  \eqref{eqn:SD1} while fixing $\frac{\Nusers}{\Nantennas} = \AntRatio$, we obtain the  achievable average ergodic rate  in the massive \ac{mimo} regime, stated in the following theorem:
\begin{theorem}
	\label{thm:SD2}
	In the massive \ac{mimo} regime, the following average ergodic rate is achievable when the \acp{bs} decode the intercell interference:   
	\begin{equation}
\RSD{} \triangleq \mathop{\lim}\limits_{\stackrel{\Nantennas \rightarrow \infty}{\frac{\Nusers}{\Nantennas} = \AntRatio}}\RSD{\Nantennas} 
	= \frac{\Tdata}{\Tcoh}\cdot\frac{1}{\Ncells  \cdot \AntRatio}   \mathop{\min}\limits_{k \in \NcellsSet} \MPFunc\left(  \AvarUp, \AntRatio\right),
	\label{eqn:SD2} 
	\end{equation}
	where $\AvarUp$ and $\MPFunc\left( \cdot, \cdot\right)$ are defined in  \eqref{eqn:AIANkDef1} and \eqref{eqn:MarPast1}, respectively.
\end{theorem}

	{\em Proof:}
%	See Appendix \ref{app:Proof5}.
%	\else 
%	\begin{IEEEproof}
	The proof follows similar arguments to the proof of Theorem \ref{thm:IAN2} and is thus omitted.
%	\end{IEEEproof}
 
%

\smallskip
\label{txt:SDExplain} 
The minimization over the cells in \eqref{eqn:SD1}-\eqref{eqn:SD2} follows since each \ac{bs}  decodes the signals of all the \acp{ut} in the network, thus the \acp{ut} have to transmit at a rate which allows their message to be reliably decoded by all \acp{bs}. Consequently, unlike the \ac{se} of treating interference as noise stated in Thm. \ref{thm:IAN2}, which always upper-bounds the \ac{se} of separate linear decoding,  simultaneous decoding can be outperformed by separate linear decoding, especially in scenarios where the intercell interference is weak. This behavior is also observed in the numerical study in Section \ref{sec:Simulations}, where it is also demonstrated that simultaneous decoding is most beneficial when the intercell interference is dominant, in agreement with its optimality for two-user Gaussian interference channels \cite[Ch. 6.4.2]{ElGamal:11}.

%-----------------------------------
%	Time Division
%----------------------------------- 
\subsection{Scheme 3 - Time Division}
\label{subsec:TD} 
Another approach is to eliminate the intercell interference by letting the \acp{ut} of different cells transmit at different time intervals. Here, the data transmission phase is divided into $\Ncells$ distinct intervals, each consisting of  $\myZeta_k \cdot \Tdata$ symbols, where $\sum\limits_{k=1}^{\Ncells} \myZeta_k = 1$. 
Unlike the schemes discussed in Subsections \ref{subsec:IAN}--\ref{subsec:SD}, this method is not a decoding scheme, but rather a method to convert the massive \ac{mimo} network into a set of non-interfering \acp{mac}.
The motivation for this approach stems from the fact that, in some scenarios, neither of the previous approaches, i.e., treating the intercell interference as noise or decoding it,	can lead to good results, and it may be preferable to cancel the intercell interference by boosting orthogonality. The drawback is that each cell now utilizes only a portion of the data transmission phase.
We note that this scheme requires a basic level of cooperation between the cells, as the \acp{ut} of different cells know not to transmit at the same time. Nonetheless, this is not the standard notation of cooperation as in \cite[Ch. 1.4]{ElGamal:11}, in the sense that no cooperative encoding or decoding is carried out, as only a basic level of centralized network control is required to allocate the time intervals between the cells.

Since each \ac{ut} in the $k$-th cell transmits in only $\myZeta_k$ of the data transmission phase, it can transmit at power of $1 / \myZeta_k$ instead of unit power, while maintaining an average unit transmission power over the  transmission phase. Consequently, the transmitted signal in the $k$-th cell during the $k$-th transmission interval is given by $\myZeta_k^{-\frac{1}{2}} \myX_k[i]$, and the corresponding channel output is 
\begin{equation}
\myY_k[i] = \hat{\Gmat}_{k,k} \myZeta_k^{-\frac{1}{2}} \myX_k[i] + \tilde{\Gmat}_{k,k} \myZeta_k^{-\frac{1}{2}} \myX_k[i] + \myW_k[i]. 
\label{eqn:TDModel1}
\end{equation}   
As no intercell interference is present, the equivalent noise is 
$\myV\TD{k}[i] \triangleq \tilde{\Gmat}_{k,k} \myZeta_k^{-\frac{1}{2}} \myX_k[i] + \myW_k[i]$, and the received signal during the $k$-th transmission interval can be written as 
\begin{equation}
\label{eqn:TDModel2}
\myY_k[i] = \hat{\Gmat}_{k,k} \myZeta_k^{-\frac{1}{2}} \myX_k[i] + \myV\TD{k}[i]. 
\end{equation}
To formulate the  \ac{se} of this scheme, we define the $\Nusers \times \Nusers$ random diagonal matrix  
\begin{equation}
\label{eqn:AggMatDefTD}
\AmatTD\!\left(  \myZeta_k\right) \triangleq  \frac{1}{{\rm Tr}\big(( \myI_{\Nusers}\! -\! \Bmat_{k,k} ) \Dmat_{k,k}^2\big)\! + \!\myZeta_k \cdot \SigW} \Bmat_{k,k}\Dmat_{k,k}^{2}. 
\end{equation}
The \ac{se} of the proposed scheme  for a finite $\Nantennas$ is stated in the following proposition:
\begin{proposition}
	\label{lem:TD1}
	When the data transmission phase is divided into $\Ncells$ distinct intervals partitions via $\{\myZeta_k\}_{k \in \NcellsSet}$, the following average ergodic rate is achievable: 
	\begin{align}
	&\hspace{-0.2cm}\RTD{\Nantennas} \left( \{\myZeta_k\}_{k \in \NcellsSet}\right)  \notag \\
	&\hspace{-0.2cm}= \frac{\Tdata}{\Tcoh}\cdot\frac{1}{\Ncells \cdot \Nusers}\sum\limits_{k =1}^{\Ncells} \myZeta_k
	\cdot
	\E \left\{\log \left|\myI_{\Nantennas} \!+ \!\UHmat \AmatTD\!\left(  \myZeta_k\right) \UHmat^H  \right|  \right\},  
	\label{eqn:TD1}
	\end{align}
	where the expectations are carried out with respect to the random matrices $\UHmat$ and $\{\AmatTD\}_{k \in \NcellsSet}$. 
\end{proposition}

{\em Proof:}
See Appendix \ref{app:Proof6}.

\smallskip
Next, we use Proposition \ref{lem:TD1} to characterize the  achievable average ergodic rate in the massive \ac{mimo} regime. To that aim,
define the set of \acp{rv} $\{A\TD{k}\}_{k \in \NcellsSet}$ such that 
\begin{equation}
\AvarTD \triangleq \frac{\bcoeff_{k,k,1}\dcoeff_{k,k,1}^{2}}{\AntRatio \cdot  \E\{\left(1 - \bcoeff_{k,k,1} \right)\dcoeff_{k,k,1}^2 \} }.
\label{eqn:ATDkDef} 
\end{equation} 
Letting  $\Nantennas \rightarrow \infty$ in  \eqref{eqn:TD1} while fixing $\frac{\Nusers}{\Nantennas} = \AntRatio$, we obtain the  achievable average ergodic rate  in the massive \ac{mimo} regime, stated in the following theorem:
\begin{theorem}
	\label{thm:TD2}
	In the massive \ac{mimo} regime, the following average ergodic rate is achievable when the data transmission phase is divided into $\Ncells$ distinct interval  via $\{\myZeta_k\}_{k \in \NcellsSet}$:  
	\vspace{-0.1cm}
	\begin{equation}
\mathop{\lim}\limits_{\stackrel{\Nantennas \rightarrow \infty}{\frac{\Nusers}{\Nantennas} = \AntRatio}}\RTD{\Nantennas}  \left( \{\myZeta_k\}_{k \in \NcellsSet}\right)
	= \frac{\Tdata}{\Tcoh}\cdot\frac{1}{\Ncells  \cdot \AntRatio}  \sum\limits_{k =1}^{\Ncells} \myZeta_k\cdot \MPFunc\left( \AvarTD ,\AntRatio \right).
	\label{eqn:TD2}
	\vspace{-0.1cm}
	\end{equation}	
\end{theorem}

{\em Proof:}
	The proof follows similar arguments to the proof of Theorem \ref{thm:IAN2} and is thus omitted.

\smallskip
Since  for each $k \in \NcellsSet$, the non-negative real-valued  $\MPFunc\left(  \AvarTD,\AntRatio \right)$ does not depend on the partitions  $\{\myZeta_k\}_{k \in \NcellsSet}$, the set of partitions which maximizes \eqref{eqn:TD2} is obtained using the Cauchy-Schwartz inequality, resulting in the following corollary:
\begin{corollary}
	\label{cor:TDOpt}
	The achievable average ergodic rate when the transmission phase is divided into $\Ncells$ intervals in the massive \ac{mimo} regime \eqref{eqn:TD2} is maximized by setting
%	\vspace{-0.2cm}
%	\begin{equation}
%	\label{eqn:TDOpt1}
	$\myZeta_k^{\rm o} = \frac{\MPFunc\left(  \AvarTD ,\AntRatio \right)}{   \sum\limits_{l=1}^{\Ncells} \MPFunc\left(  \AvarTD[l] ,\AntRatio \right)}$, for all $k \in \NcellsSet$,
%	\vspace{-0.2cm}
%	\end{equation}
	and the resulting achievable average  ergodic rate is given by
	\begin{align} 
\RTD{} &\triangleq  \mathop{\lim}\limits_{\stackrel{\Nantennas \rightarrow \infty}{\frac{\Nusers}{\Nantennas} = \AntRatio}}\RTD{\Nantennas} \left( \{\myZeta_k^{\rm o}\}_{k \in \NcellsSet}\right) \notag \\  
	&= \frac{\Tdata}{\Tcoh}\cdot\frac{1}{\Ncells  \cdot \AntRatio} \cdot \frac{ \sum\limits_{k =1}^{\Ncells} \MPFunc^2\left( \AvarTD ,\AntRatio \right)}{\sum\limits_{k =1}^{\Ncells} \MPFunc\left( \AvarTD ,\AntRatio \right)}.
	\label{eqn:TDOpt2} 
	\end{align}		
\end{corollary}

%-----------------------------------
%	Illustrative Example
%-----------------------------------
\vspace{-0.2cm}
\subsection{Illustrative Example}
\label{subsec:Example}
\vspace{-0.1cm}
In order to analytically illustrate the relationships between \acp{se} of the schemes discussed in the previous subsections, we consider, as an example, a massive \ac{mimo} network consisting of $\Ncells = 2$ cells in the high \ac{snr} regime, i.e., $\SigW \rightarrow 0$. To properly formulate this example, let $\Xrv \in \big[\Xmin, \Xmax\big]$ and $\Yrv \in \big[\Ymin, \Ymax\big]$ be mutually independent \acp{rv} of finite support, where $0 < \Xmin < \Xmax$ and $0 < \Ymin < \Ymax$. For every $m \in \NusersSet$, the attenuation coefficients are distributed via $\dcoeff_{k,l,m}^2 \Dist \Xrv$ for $k=l$ and $\dcoeff_{k,l,m}^2 \Dist \Yrv$ for $k \ne l$. 
In particular, we consider two extreme interference profiles:
%
%\begin{enumerate}
%	\item 
	$1)$ $\Ymax \ll \Xmin$ - this case represents weak intercell interference.
%	\item  
	$2)$ $\Xmax \ll \Ymin$  - this case corresponds to dominant intercell interference.
%\end{enumerate}
Note that these interference profiles resemble the weak interference regime and the strong interference regime, respectively, traditionally defined for the two-user Gaussian non-fading interference channel \cite[Ch. 6.4]{ElGamal:11}.
The relationships between the asymptotic \acp{se} in Theorems \ref{thm:IAN2}-\ref{thm:SD2} and Corollary \ref{cor:TDOpt} for these scenarios are stated in the following proposition:
\begin{proposition}
	\label{pro:Example}
	When $\Ymax \ll \Xmin$, the asymptotic \acp{se}  satisfy $\RTD{} \approx \RSD{}$ and $\RIAN{} \approx 2\RSD{}$,
	while for $\Xmax \ll \Ymin$, these \acp{se} satisfy $\RIAN{} \approx 0$ and	$\RTD{} < \RSD{}$.
\end{proposition}

{\em Proof:}
See Appendix \ref{app:Proof9}.

\smallskip
Proposition \ref{pro:Example} agrees with the theoretical results for the two-user Gaussian interference, for which it is known that treating interference as noise is optimal in the weak interference regime, while simultaneous decoding is optimal in the strong interference regime \cite[Ch. 6.4]{ElGamal:11}. 
In the numerical study in Section \ref{sec:Simulations} we demonstrate that time division can contribute to increasing the \ac{se} when the interference is not too weak and not too dominant. 
Furthermore, 
the proposition implies that in the weak interference regime, the \ac{se} of treating interference as noise is larger by a factor of approximately $\Ncells$ compared to simultaneous decoding and time division.
Since  $\RIAN{} \approx 0$  when the intercell interference is dominant, Proposition \ref{pro:Example} indicates that any approach that is based on treating intercell interference as noise, including the standard separate linear decoding approach, is expected to result in negligible \ac{se} when the intercell interference is dominant, and cannot approach the fundamental rate limits in such scenarios.

%-----------------------------------
%	Optimized Scheme
%-----------------------------------
\vspace{-0.25cm}
\subsection{Optimized Scheme}
\label{subsec:Optimize} 
\vspace{-0.1cm}
To benefit from the advantages of  Schemes 1--3  we propose a method which combines them in order to optimize the overall \ac{se}. 
Generally speaking, the proposed optimized approach allows time division as in Scheme 3 by partitioning the transmission phase where only some of the cells in the network are active at each partition, and combines the decoding schemes 1--2 by allowing each \ac{bs} to jointly decode some of the intercell interference, and treat the rest as noise. 
Specifically, we let the transmission phase $\Tdata$ be divided into $\Nparts \le \Ncells$ distinct intervals, with the $q$-th interval consisting of $\myZeta_q \cdot\Tdata$ symbols, $q \in \{1,2,\ldots, \Nparts\}$, where $\sum\limits_{q=1}^{\Nparts}\myZeta_q = 1$.  
We let $\Part{q}$ denote the set of active cells during the $q$-th interval, such that $\bigcup\limits_{q=1}^{\Nparts} \Part{q} = \NcellsSet$ and $\Part{q_1} \bigcap \Part{q_2} = \EmpSet$ for every $q_1 \ne q_2$. During the $q$-th interval, only the \acp{ut} belonging to the set of active cells $\Part{q}$ are allowed to transmit\footnote{{\label{ftn:Opimality}We note that the \ac{se} can further optimized by allowing the cells to be active on more than one transmission interval, namely, by removing the restriction $\mathcal{I}_{q_1} \cap \mathcal{I}_{q_2} = \EmpSet$ for each $q_1 \ne q_2$. However, as the purpose of the scheme is to show that the \ac{se} can be optimized by properly combining schemes 1-3, we defer this generalization to future exploration.}}, and each \ac{ut} transmits at power of $1 / \myZeta_q$ instead of unit power. 
Next, we divide the active cells in each interval $q$ into $\Nclusts \le \left|\Part{q} \right|$ distinct non-empty clusters, denoted $\{\Clust \}_{s = 1}^{\Nclusts}$, such that  $\bigcup\limits_{s=1}^{\Nclusts} \Clust = \Part{q}$. 
During the $q$-th interval, each \ac{bs} $k \in \Clust$ treats the intercell interference from the cells in the set $\NegClust \triangleq \Part{q} \setminus \Clust$ as noise, and decodes the signals of the \acp{ut} of the cells $\Clust$. 

In the following we characterize the \ac{se} for a fixed setting of clusters $\{\Clust \}_{s = 1, q=1}^{\Nclusts, \Nparts}$ in the massive \ac{mimo} regime.  
 The received signal at the $k$-th \ac{bs} %during data transmission 
 \eqref{eqn:Channel_Data1}, $k \in \Clust$, can be written as\footnote{Since the sets $\{\Clust \}_{s = 1, q=1}^{\Nclusts, \Nparts}$ are distinct and span the set of cells $\NcellsSet$, the values of the partition index $q$ and the cluster index $s$ are uniquely determined by the cell index $k$, i.e., $q=q(k)$ and $s=s(k)$. For notational simplicity, we omit the cell index $k$.} 
 \begin{align}
 \myY_k[i] &=  \sum\limits_{l\in \Clust}\hat{\Gmat}_{k,l}\myZeta_q^{-\frac{1}{2}} \myX_l[i] + \sum\limits_{l\in \Clust}\tilde{\Gmat}_{k,l}\myZeta_q^{-\frac{1}{2}}\myX_l[i] \notag \\ 
 &+   \sum\limits_{l\in \NegClust}{\Gmat}_{k,l}\myZeta_q^{-\frac{1}{2}} \myX_l[i] + \myW_k[i]. 
 \label{eqn:OSModel1}
 \end{align} 
 When decoding the intercell interference from the cells belonging to the set $\Clust$ along with the data, the equivalent noise is 
 $\myV\OS{k}[i] \triangleq \sum\limits_{l\in \Clust}\tilde{\Gmat}_{k,l}\myZeta_q^{-\frac{1}{2}}\myX_l[i]+   \sum\limits_{l\in \NegClust}{\Gmat}_{k,l}\myZeta_q^{-\frac{1}{2}} \myX_l[i]  + \myW_k[i]$, and the received signal can be written as 
 \begin{align}
 \myY_k[i] &= \sum\limits_{l\in \Clust}\hat{\Gmat}_{k,l}\myZeta_q^{-\frac{1}{2}} \myX_l[i] + \myV\OS{k}[i] 
 \notag \\
 &
 = \myZeta_q^{-\frac{1}{2}} \hat{\Gmat}_{k,k}{\Dmat}_{k,k}^{-2}\sum\limits_{l\in \Clust}{\Dmat}_{k,l}^2\myX_l[i] + \myV\OS{k}[i].
 \label{eqn:OSModel2} 
 \end{align}
The representation \eqref{eqn:OSModel2} facilitates the characterization of the \ac{se}. By defining the scalar  \acp{rv} 
\vspace{-0.1cm}
%\begin{subequations}
%	\label{eqn:AOSkDef}
	\begin{equation*}
	\AvarUpOS\!\left(\{\Clust \}_{s = 1}^{\Nclusts} \right)  \!\triangleq\! 
	%\left( {\AntRatio  \sum\limits_{l=1}^{\Ncells} \E\{\left(1 - \bcoeff_{k,l,1} \right)\dcoeff_{k,l,1}^2 \} }\right)^{-1}\!\!\!\! {\bcoeff_{k,k,1}\dcoeff_{k,k,1}^2}.
	%
	\frac{\bcoeff_{k,k,1}\dcoeff_{k,k,1}^{-2} \sum\limits_{l\in \Part{q}}\dcoeff_{k,l,1}^{4}} {\AntRatio  \sum\limits_{l\in \Part{q}} \E\{\left(1 \!-\! \bcoeff_{k,l,1} \right)\dcoeff_{k,l,1}^2 \} };
%	\vspace{-0.1cm}
%	\label{eqn:AOSkDef1}
	\end{equation*} 
	\begin{equation*}
\, 
	\AvarDnOS\!\left(\{\Clust \}_{s = 1}^{\Nclusts} \right)\!\triangleq\! 
	%\left( {\AntRatio  \sum\limits_{l=1}^{\Ncells} \E\{\left(1 - \bcoeff_{k,l,1} \right)\dcoeff_{k,l,1}^2 \} }\right)^{-1}\!\!\!\! {\bcoeff_{k,k,1}\dcoeff_{k,k,1}^2}.
	%
	\frac{\bcoeff_{k,k,1}\dcoeff_{k,k,1}^{-2} \sum\limits_{l\in \NegClust}\dcoeff_{k,l,1}^{4}} {\AntRatio  \sum\limits_{l\in \Part{q}} \E\{\left(1\! -\! \bcoeff_{k,l,1} \right)\dcoeff_{k,l,1}^2 \} },
	\vspace{-0.1cm}
	\label{eqn:AOSkDef2}
	\end{equation*}
	and the deterministic quantity  
	\begin{align*}
	\myZq{q} \left(\{\Clust \}_{s = 1}^{\Nclusts} \right)  \triangleq \sum\limits_{s=1}^{\Nclusts}&\mathop{\min}\limits_{k \in \Clust}\bigg(  \MPFunc\left( \AvarUpOS\left(\{\Clust \}_{s = 1}^{\Nclusts} \right) ,\AntRatio \right) \notag \\
	& -  \MPFunc\left( \AvarDnOS\left(\{\Clust \}_{s = 1}^{\Nclusts} \right) ,\AntRatio \right)\bigg),
%	\label{eqn:AOSkDef3} 
	\end{align*}
%\end{subequations}
we obtain the  \ac{se}  in the massive \ac{mimo} regime, as stated in the following theorem:
\begin{theorem}
	\label{thm:OS2}
	In the massive \ac{mimo} regime, the following average ergodic rate is achievable for a fixed setting of clusters $\{\Clust \}_{s = 1, q=1}^{\Nclusts, \Nparts}$ and partitions $\{\myZeta_q\}_{q=1}^{\Nparts}$:   
	\begin{align}
	&\ROS{}  \left(\{\Clust \}_{s = 1, q=1}^{\Nclusts, \Nparts}, \{\myZeta_q\}_{q=1}^{\Nparts}  \right) \notag \\
	&= \frac{\Tdata}{\Tcoh}\cdot\frac{1}{\Ncells  \cdot \AntRatio}  \sum\limits_{q =1}^{\Nparts} \myZeta_q \cdot \myZq{q} \left(\{\Clust \}_{s = 1}^{\Nclusts} \right).
	\label{eqn:OS2} 
	\end{align}	
\end{theorem}

{\em Proof:}
See Appendix \ref{app:Proof8}. 

\smallskip
Note that Theorem \ref{thm:OS2} specializes Theorems \ref{thm:IAN2}--\ref{thm:TD2} by properly setting  $\{\Clust \}_{s = 1, q=1}^{\Nclusts, \Nparts}$. In particular:
\begin{itemize}
	\item When $\Nparts = 1$ (i.e., $\Part{1} = \NcellsSet$), and $\Nclusts = \Ncells$ (i.e., each cluster $\Clust$ contains only one cell), then  $\AvarUpOS$ and $\AvarDnOS$ coincide with $\AvarUp$ and $\AvarDn$, respectively, and  \eqref{eqn:OS2} reduces to \eqref{eqn:IAN2}.
	\item For $\Nparts = 1$ (i.e., $\Part{1} = \NcellsSet$), and $\Nclusts = 1$ (i.e., a single cluster which contains all the cells in the network, $\Clust = \NcellsSet$), we have that  $\AvarUpOS$  coincides with $\AvarUp$, while $\AvarDnOS$ is zero with probability $1$, and thus \eqref{eqn:OS2} specializes to \eqref{eqn:SD2}.
	\item By setting $\Nparts = \Ncells$ (i.e., only one active cell in each partition),	we have that  $\AvarUpOS$  coincides with $\AvarTD$, while $\AvarDnOS$ is zero with probability $1$, and thus \eqref{eqn:OS2} reduces to \eqref{eqn:TD2}.
\end{itemize}
Furthermore, as in Corollary \ref{cor:TDOpt}, the set of partitions $\{\myZeta_q\}_{q=1}^{\Nparts}$ which maximizes \eqref{eqn:OS2} for a fixed set of clusters $\{\Clust \}_{s = 1, q=1}^{\Nclusts, \Nparts}$ can be explicitly obtained using the Causchy-Schwartz inequality as 
	\begin{equation}
	\label{eqn:OSOpt1}
	\myZeta_q^{\rm o}\left(\{\Clust \}_{s = 1}^{\Nclusts} \right) = \frac{\myZq{q}\left(\{\Clust \}_{s = 1}^{\Nclusts} \right)}
	{   \sum\limits_{q'=1}^{\Nparts} \myZq{q'}\left(\{\Clust \}_{s = 1}^{\Nclusts} \right)}. 
	\end{equation}
Finally, we combine Theorem \ref{thm:OS2} and \eqref{eqn:OSOpt1} to formulate an optimization problem whose solution is the maximal \ac{se} by any combination of the schemes 1--3, stated in the following corollary:
\begin{corollary}
	\label{cor:OS3}
	In the massive \ac{mimo} regime, the following average ergodic rate is achievable: 
	\begin{align}
	r\OS{\max} = &\mathop{\max}\limits_{\Nparts, \{\Nclusts \}_{ q=1}^{ \Nparts}, \{\Clust \}_{s = 1, q=1}^{\Nclusts, \Nparts}} \Bigg(\frac{\Tdata}{\Tcoh}\cdot\frac{1}{\Ncells  \cdot \AntRatio} \notag \\
	&\qquad\qquad\cdot  	
	\frac{ \sum\limits_{q =1}^{\Nparts}\left( \myZq{q}\left(\{\Clust \}_{s = 1}^{\Nclusts} \right)\right)^2 }
	 { \sum\limits_{q =1}^{\Nparts} \myZq{q}\left(\{\Clust \}_{s = 1}^{\Nclusts} \right) } \Bigg),
	 \label{eqn:OS3}
	\end{align}
	where $1 \le \Nparts \le \Ncells$, and $ \{\Clust \}_{s = 1, q=1}^{\Nclusts, \Nparts}$ are non-empty distinct sets  which span $\NcellsSet$.
\end{corollary}

The achievable average ergodic rate is given by the optimization problem in \eqref{eqn:OS3}, where the parameters over which the optimization is carried out are the number of partitions $\Nparts$, the number of clusters in each partitions  $ \{\Nclusts \}_{ q=1}^{ \Nparts}$, and the cells allocated to each cluster $\{\Clust \}_{s = 1, q=1}^{\Nclusts, \Nparts}$. Thus, the optimization is carried out over a finite set, and can be solved by searching over all possible combinations of $\Nparts$, $\{\Nclusts \}_{ q=1}^{ \Nparts}$, and $\{\Clust \}_{s = 1, q=1}^{\Nclusts, \Nparts}$. 
Note that  \eqref{eqn:OS3} considers only the overall \ac{se}. Other parameters which may be of interest in practical networks, such as fairness \cite{Bethanabhotla:16}, can be accounted for by introducing additional constraints on the sets of clusters and partitions.
While solving \eqref{eqn:OS3} may be computationally difficult, especially for a large number of cells, its solution is expected to provide an indication of the underlying fundamental performance limits of  uplink  massive \ac{mimo} systems. 
In particular, the gain of the optimized scheme stems from the fact that it combines schemes 1-3, allowing each \ac{bs} to decode the signals from some cells, treat the signals from other cells as noise, while canceling the interference from the rest of the cells via time-division. Therefore, its gain over schemes 1-3 is most notable in scenarios where the interference profiles vary significantly between cells, and neither of the aforementioned approaches is optimal, as also demonstrated in the numerical study detailed in Section \ref{sec:Simulations}.

%----------------------------------------------------------------------------------------
%	Numerical Results and Discussion
%---------------------------------------------------------------------------------------- 
\section{Numerical Results and Discussion}
\label{sec:Simulations} 
In this section we evaluate the achievable average ergodic rates of massive \ac{mimo} networks using the schemes discussed in Section \ref{sec:Rates} in a simulations study, consisting of two parts: 
First, in Subsection \ref{subsec:MassiveSim}
 we numerically evaluate the number of \ac{bs} antennas which can be considered as the massive \ac{mimo} regime, i.e., for which values of  $\Nantennas$, our asymptotic analysis in Theorems \ref{thm:IAN2}--\ref{thm:TD2} accurately characterizes the achievable average ergodic rates.  
In the second part  
in Subsection \ref{subsec:SESim} 
 we compare the \acp{se} of the schemes detailed in Section \ref{sec:Rates} to the  rates achievable using standard separate linear decoding in the massive \ac{mimo} regime. 

We consider a network  
consisting of $\Ncells = 5$ cells. The coherence duration is $\Tcoh = 1000$ symbols.  For each Monte Carlo simulation,  
the attenuation coefficients are generated as $\dcoeff_{k,l,m} = \frac{Z_{k,l,m}}{C_{k,l,m}^2}$, where $\{Z_{k,l,m}\}$ are the shadow fading coefficients, independently randomized from a log-normal distribution with standard deviation of $ 8$ dB, and  $\{C_{k,l,m}\}$ represent the range between the $m$-th \ac{ut} of the $l$-th cell and the $k$-th \ac{bs}, $k,l \in \NcellsSet$, $m \in \NusersSet$ \cite[Sec. II-C]{Marzetta:10}. 
In the first part of our study we consider a synthetic model for $\{C_{k,l,m}\}$, which we discuss in the sequel, used to evaluate our results while directly controlling the level of intercell interference. In our final simulations study we use a realistic model which more faithfully represents cellular networks.

To formulate the synthetic model for $\{C_{k,l,m}\}$, let $(( \cdot ))_{\Ncells}$ be the modulo $\Ncells$ operator, and $\{U_{k,l,m}\}$ be i.i.d. \acp{rv} uniformly distributed over $[1,2]$. In order to capture various interference profiles, we use three different distributions for the \acp{rv} $C_{k,l,m}$: 
\begin{itemize}
	\item $C^2_{k,l,m} = e^{3 (( k-l ))_{\Ncells}}\cdot U_{k,l,m}$,  we refer to this setting as {\em weak interference}.
	\item $C^2_{k,l,m} = e^{0.25 (( k-l ))_{\Ncells}}\cdot U_{k,l,m}$,  we refer to this setting as {\em moderate interference}.
	\item $C^2_{k,l,m} = e^{-1 (( k-l ))_{\Ncells}}\cdot U_{k,l,m}$,  we refer to this setting as {\em strong interference}.
\end{itemize}  
Stochastic expectations are evaluated by averaging over $2000$ Monte Carlo simulations.  
By controlling the distribution of the distances between the \acp{ut} and the \acp{bs}, represented via the \acp{rv}  $\{C_{k,l,m}\}$, we simulate different intercell interference profiles. For example, in the weak interference setting, the \acp{ut} are significantly closer to their associated \ac{bs} than to any of the other \acp{bs}, resulting in a low level of intercell interference. In the strong interference setting, each \ac{ut} is likely to be closer to a \ac{bs} of a different cell than to the \ac{bs} of its cell, resulting in dominant intercell interference.

%-----------------------------------
%	Massive \ac{mimo} Regime Evaluation
%-----------------------------------
\subsection{Massive \ac{mimo} Regime Evaluation}
\label{subsec:MassiveSim} 
We first numerically evaluate the number of \ac{bs} antennas $\Nantennas$ for which our asymptotic \ac{se} analysis in Theorems \ref{thm:IAN2}--\ref{thm:TD2} coincide with their corresponding finite-antenna counterparts in Propositions \ref{lem:IAN1}--\ref{lem:TD1}. To that aim, we fix the number of pilot symbols  used for channel estimation to $\Tpilots = 100$,  the number of \acp{ut} in each cell to $\Nusers = 40$,  and the  \ac{snr}, defined as $1 / \SigW$, to $0$ dB. The asymptotic \acp{se} computed via  Theorems \ref{thm:IAN2}--\ref{thm:TD2} compared to the non-asymptotic \acp{se} computed via  Propositions \ref{lem:IAN1}--\ref{lem:TD1} are depicted in Figures   \ref{fig:NumEx1a} and \ref{fig:NumEx1b}  for the weak interference and for the moderate interference settings, respectively. Since the optimal time partition for the time division scheme is given in Corollary \ref{cor:TDOpt} only for the asymptotic regime, the \acp{se} of the time division scheme in Proposition \ref{lem:TD1} and Theorem \ref{thm:TD2} are computed with equal time partitions, i.e., $\myZeta_k = \Ncells^{-1}$, $\forall k \in \NcellsSet$.
 
Observing Figs.  \ref{fig:NumEx1a}--\ref{fig:NumEx1b}, we note  an excellent match between the non-asymptotic and asymptotic analysis for number of \ac{bs} antennas above $\Nantennas = 160$.
{Note that the asymptotic scheme detailed in  Subsection \ref{subsec:Optimize} essentially combines schemes 1-3, thus its asymptotic analysis also holds for such values of $\Nantennas$.} 
This indicates that the asymptotic analysis can be used to characterize the  achievable  average ergodic rates when each \ac{bs} is equipped with a large, finite number of antennas, in the order of hundreds or more \ac{bs} antennas, which is the same order as the conventional massive \ac{mimo} regime \cite{Lu:14}. 

\ifsingle
   \begin{figure}
   	\centering
   	\begin{minipage}{0.45\textwidth}
   		\centering
		\scalebox{0.48}{\includegraphics{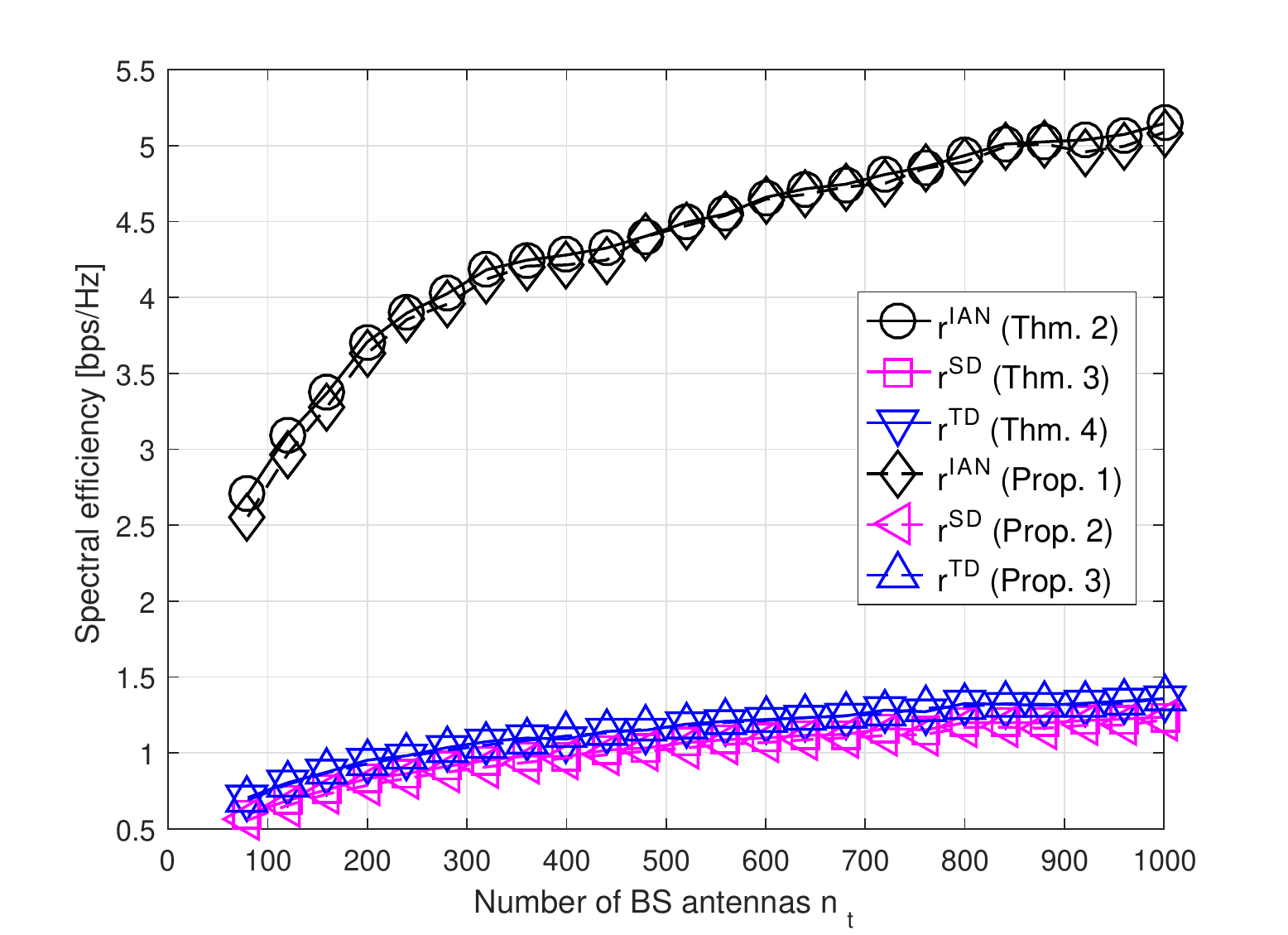}}
   		\vspace{-0.8cm}
   		\caption{Finite vs. asymptotic analysis, weak interference, $\Nusers = 40$, ${\rm SNR} = 0$ dB.}
   		\label{fig:NumEx1a}		
   	\end{minipage}
   	$\quad$
   	\begin{minipage}{0.45\textwidth}
   		\centering
		\scalebox{0.48}{\includegraphics{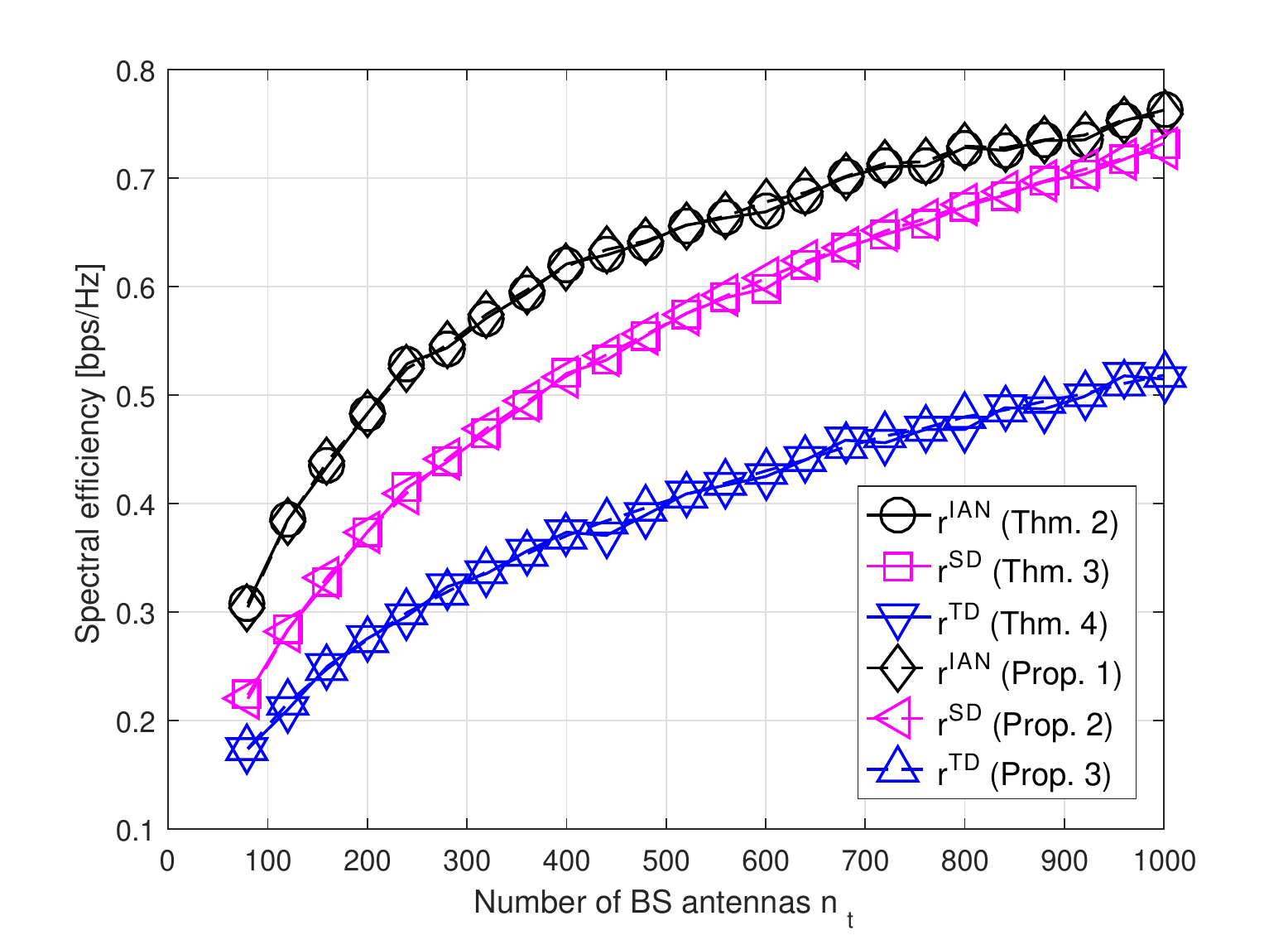}}
		\vspace{-0.8cm}
		\caption{Finite vs. asymptotic analysis, moderate interference, $\Nusers = 40$, ${\rm SNR} = 0$ dB.}
   		\label{fig:NumEx1b}
   	\end{minipage}
   	\vspace{-0.9cm}
   \end{figure}
 
\else
\begin{figure}
	\centering
	\vspace{-0.2cm}
	\includegraphics[width=7cm, height=5.2cm]{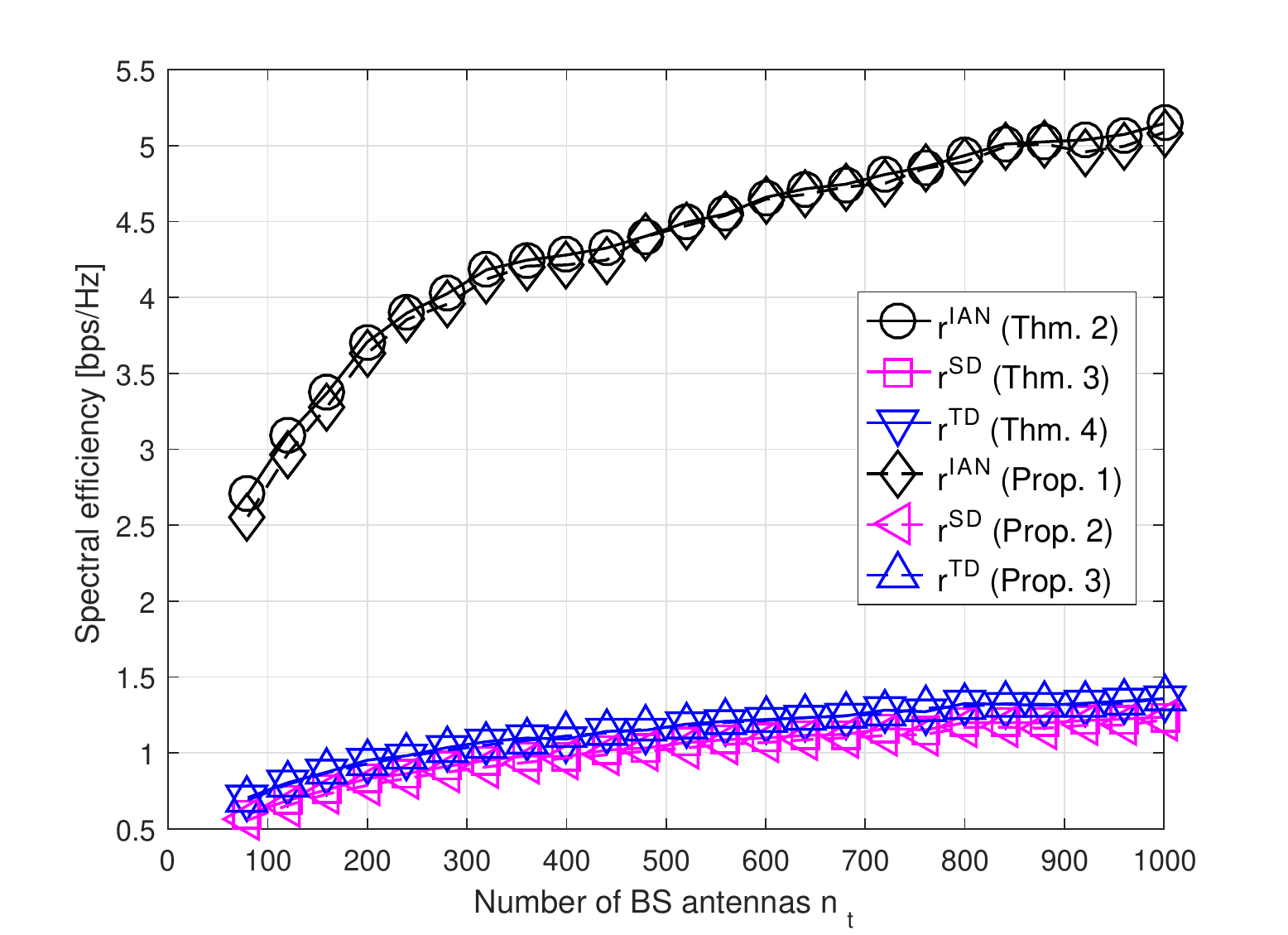}
	\vspace{-0.4cm}
	\caption{Finite vs. asymptotic analysis, weak interference, $\Nusers = 40$, ${\rm SNR} = 0$ dB.}
	\vspace{-0.2cm}
	\label{fig:NumEx1a}
\end{figure}
\begin{figure}
	\centering
	\vspace{-0.2cm}
	\includegraphics[width=7cm, height=5.2cm]{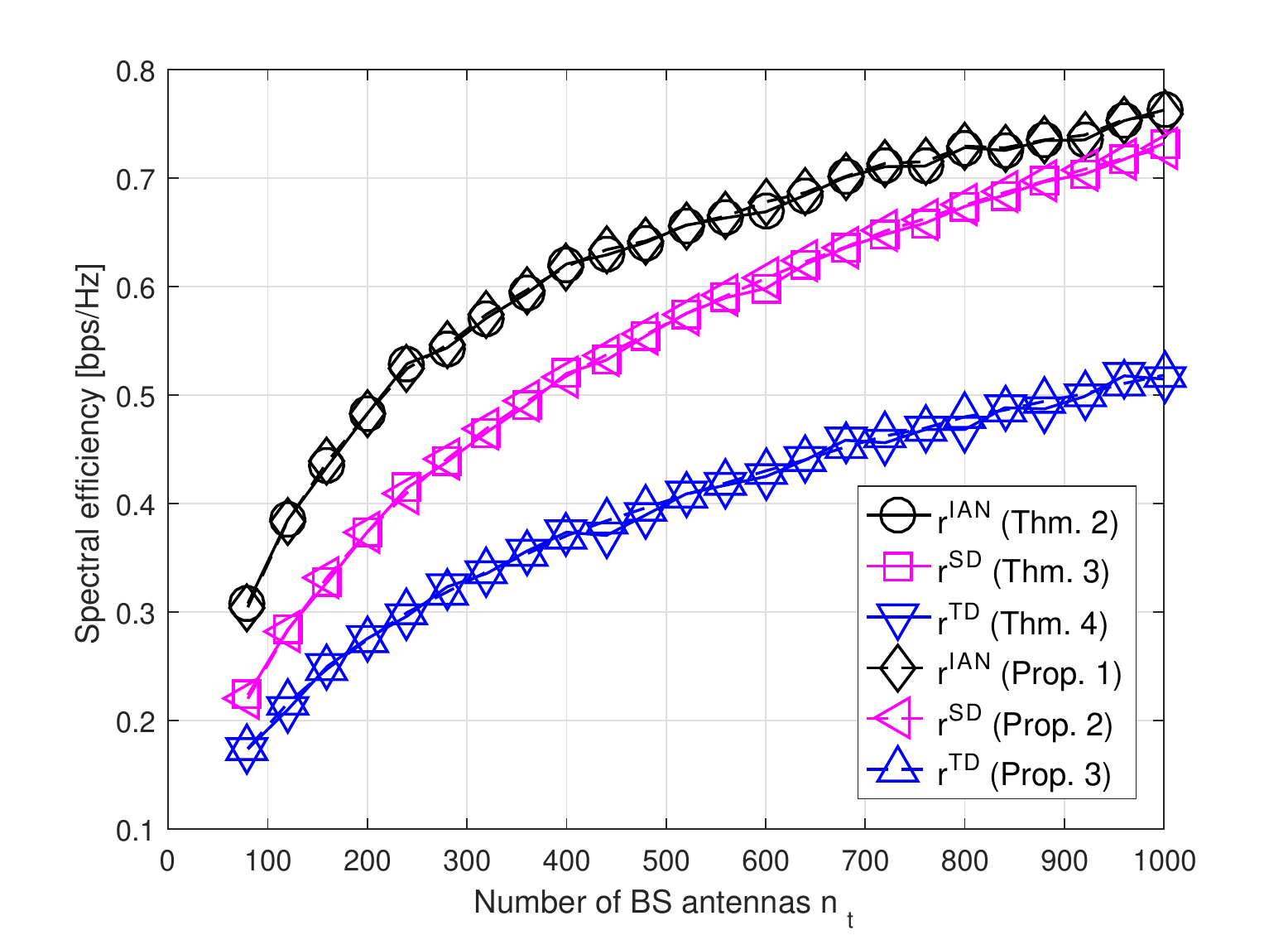}
	\vspace{-0.4cm}
	\caption{Finite vs. asymptotic analysis, moderate interference, $\Nusers = 40$, ${\rm SNR} = 0$ dB.}
	\vspace{-0.2cm}
	\label{fig:NumEx1b}
\end{figure}
\fi %-------------------------------------
%% TODO NIR divide into two figures and add non asymptotic TD results.

%-----------------------------------
%	Asymptotic \ac{se} Comparison
%-----------------------------------| 
\subsection{Asymptotic \ac{se} Comparison}
\label{subsec:SESim} 
We now compare the asymptotic \acp{se} of the schemes detailed in Section \ref{sec:Rates} to the corresponding rates achievable using separate decoding  with matched filtering and with \ac{mmse} filtering,  computed  via \eqref{eqn:SepLinRate}, where the \ac{sinr} is computed using \cite[Thm. 3]{Hoydis:13}, 
by averaging over all generated channel realizations. Here,  the number of \ac{bs} antennas is  $\Nantennas \!=\! 800$, and the number of \acp{ut} in each cell is $\Nusers \!=\! 80$. 
The achievable average ergodic rate of the time-division scheme is computed assuming optimal time partition, namely, via Corollary \ref{cor:TDOpt}. Since time division can be considered as a form of cooperation between the cells, we compute the \ac{se} of the optimized scheme twice: once with optimal time division, via Corollary \ref{cor:OS3}, and once with no time division, by maximizing the \ac{se} in Theorem \ref{thm:OS2} with  $\Nparts = 1$. 
To evaluate the \ac{se} versus \ac{snr},  $1 / \SigW$,  we fix the number of symbols  used for channel estimation to $\Tpilots = 100$, and let the \ac{snr} vary from $-30$ dB to $30$ dB. 

\begin{figure}
 	\centering
 	\begin{minipage}{0.45\textwidth}
 		\centering
 		\scalebox{0.48}{\includegraphics{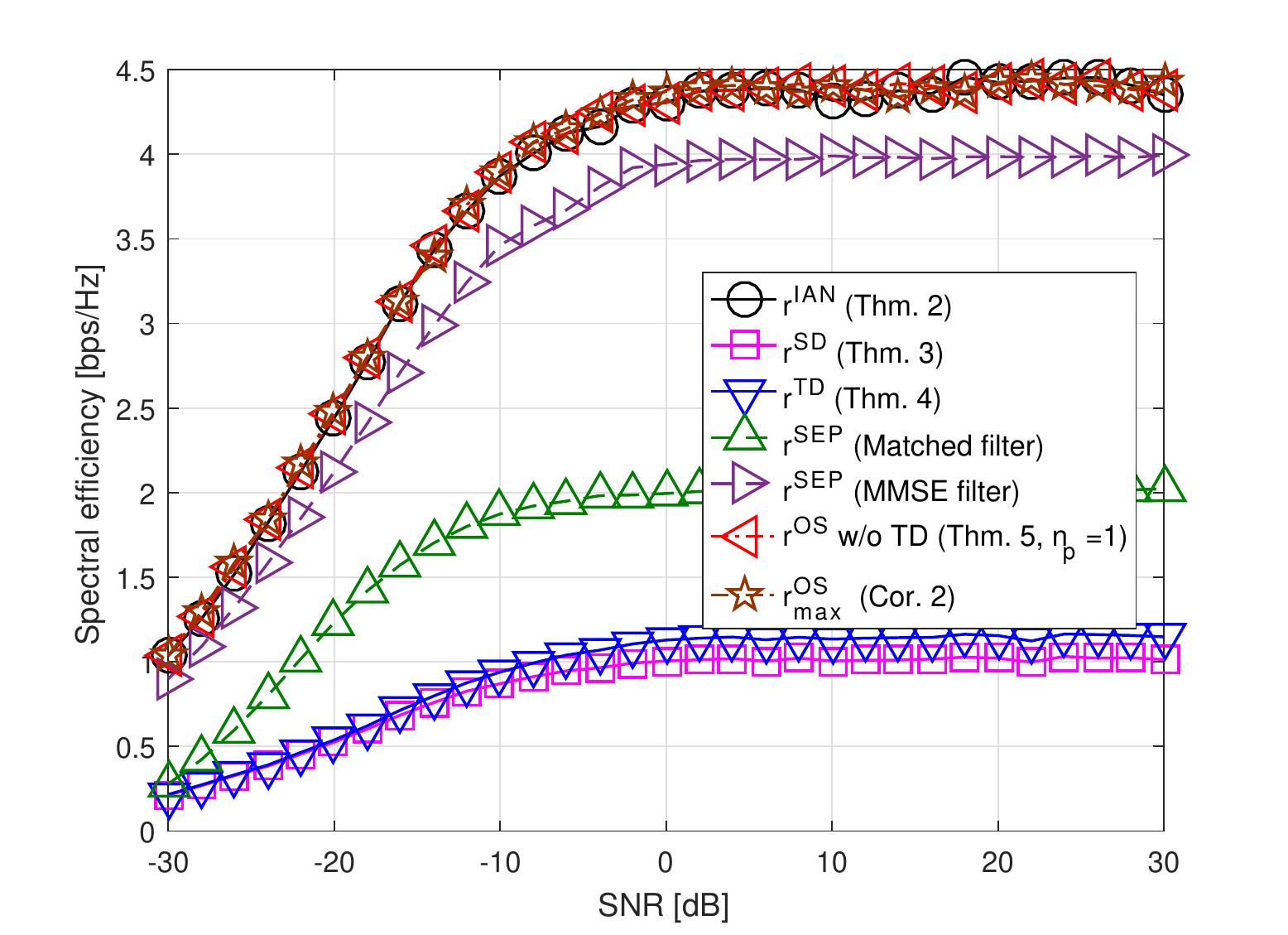}}
 		\vspace{-0.4cm}
 		\caption{\ac{se} vs. \ac{snr}, weak interference.}% $\Nusers = 80$,  $\Nantennas = 800$.}
 		\label{fig:NumEx2a}		
 	\end{minipage}
 	$\quad$
 	\begin{minipage}{0.45\textwidth}
 		\centering
 		\scalebox{0.48}{\includegraphics{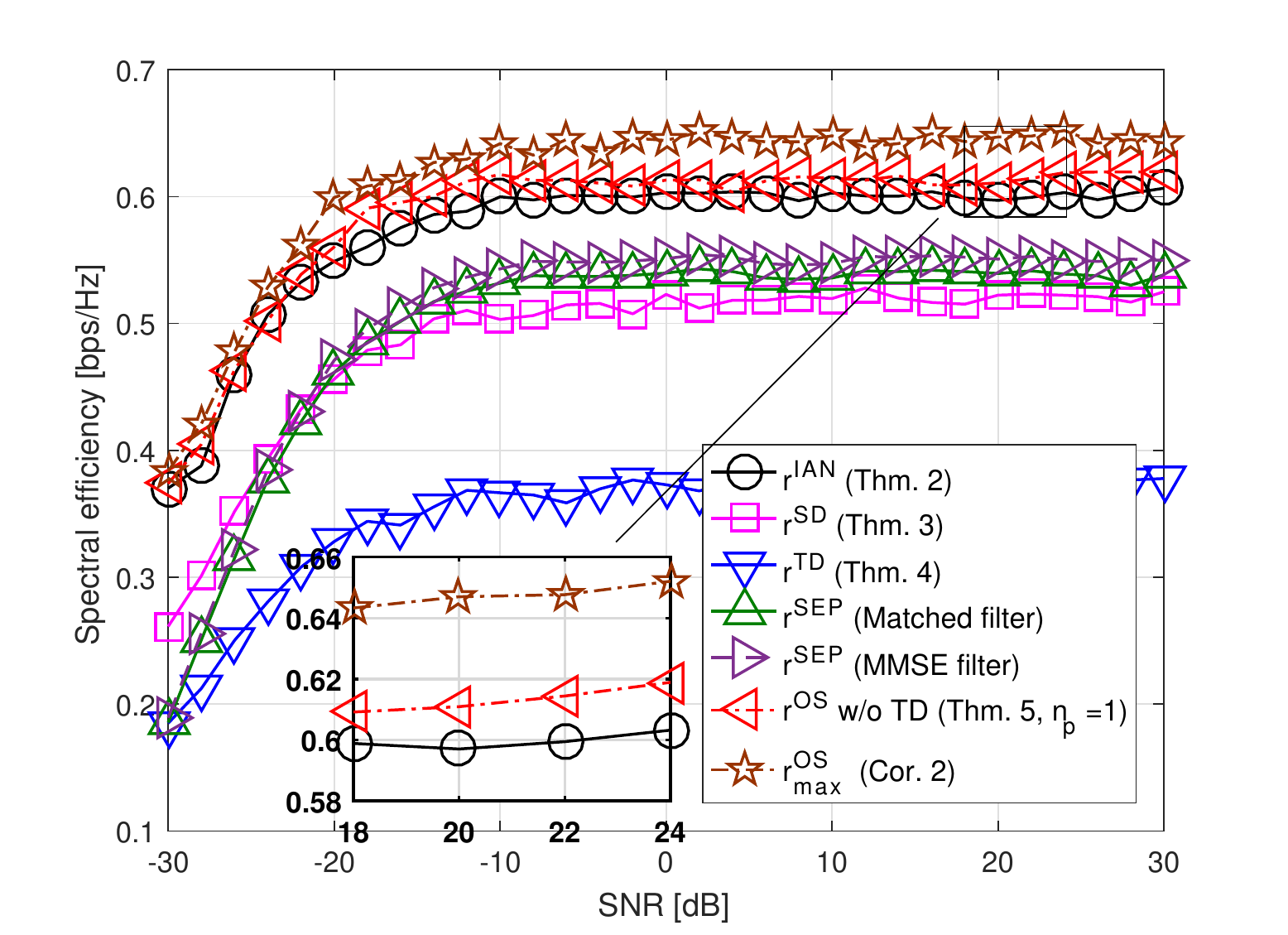}}
 		\vspace{-0.4cm}
 		\caption{\ac{se} vs. \ac{snr}, moderate interference.}% $\Nusers = 80$,  $\Nantennas = 800$.}
 		\label{fig:NumEx2b}
 	\end{minipage}
 	\vspace{-0.3cm}
 \end{figure}

\begin{figure}
 	\centering
 	\begin{minipage}{0.45\textwidth}
 		\centering
 		\scalebox{0.48}{\includegraphics{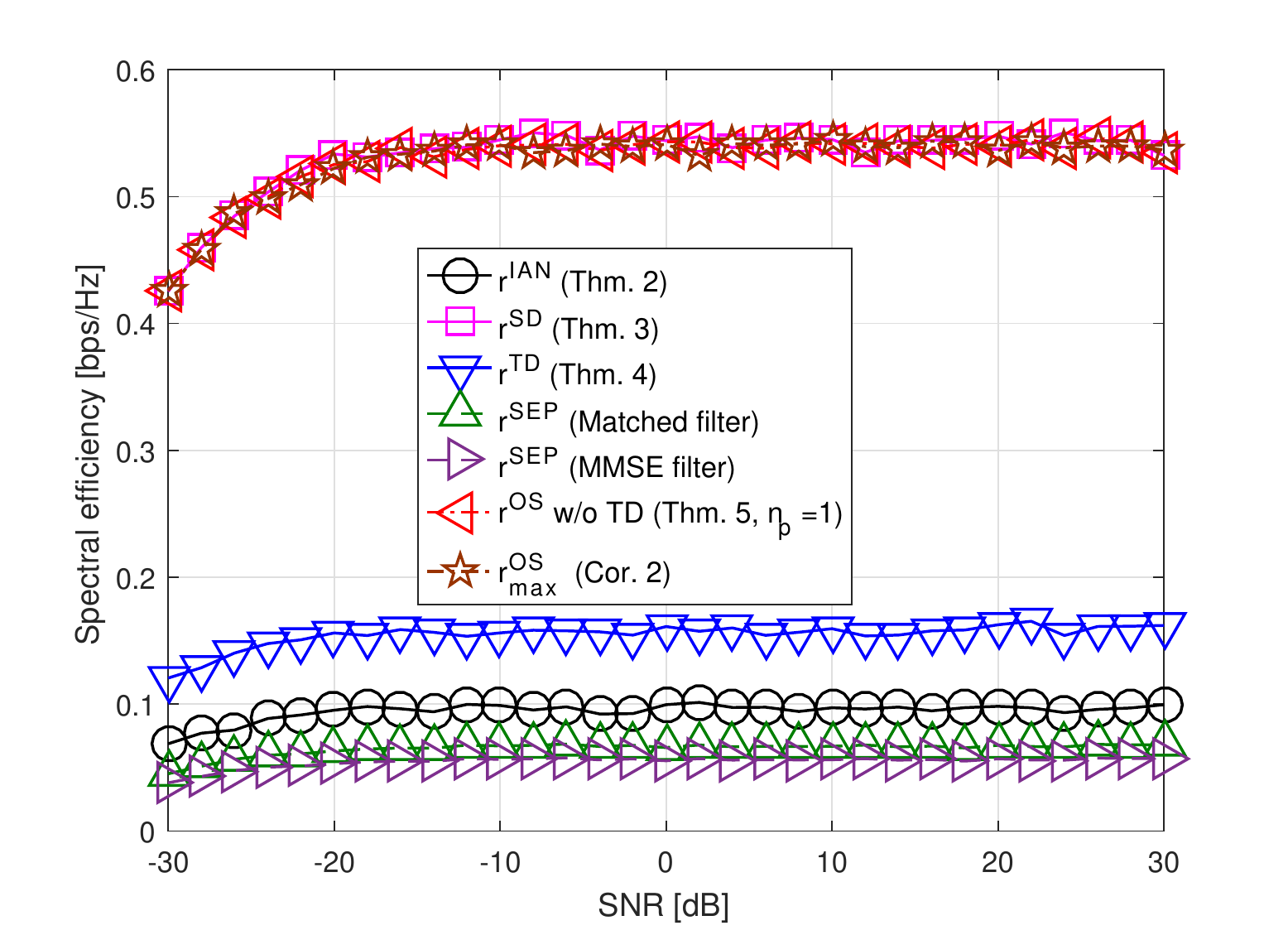}}
 		\vspace{-0.4cm}
 		\caption{\ac{se} vs. \ac{snr}, strong interference.}% $\Nusers = 80$,  $\Nantennas = 800$.}
 		\label{fig:NumEx2c}		
 	\end{minipage}
 	$\quad$
 	\begin{minipage}{0.45\textwidth}
 		\centering
 		\scalebox{0.48}{\includegraphics{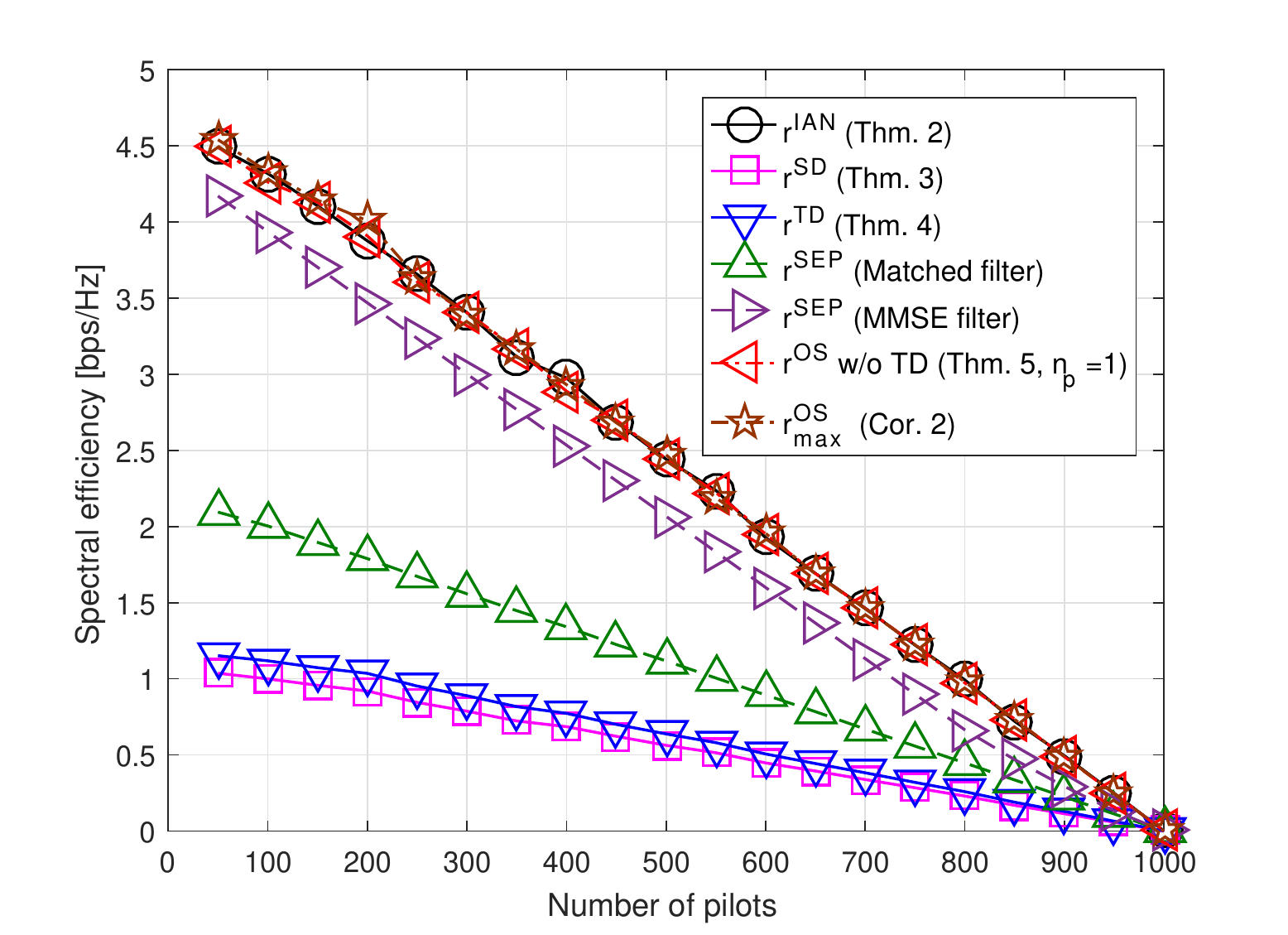}}
 		\vspace{-0.4cm}
 		\caption{\ac{se} vs. number of pilots, weak interference.}% $\Nusers = 80$,  $\Nantennas = 800$.}
 		\label{fig:NumEx3a}
 	\end{minipage}
 	\vspace{-0.7cm}
 \end{figure}

The results for the weak interference, moderate interference, and strong interference settings the are depicted in Figs. \ref{fig:NumEx2a}, \ref{fig:NumEx2b}, and \ref{fig:NumEx2c}, respectively. 
As expected, the optimized scheme obtains the highest \ac{se} in each setting over the entire \ac{snr} range,  providing an indication on the true fundamental limits of massive \ac{mimo} systems. 
Furthermore, we observe in Fig. \ref{fig:NumEx2a} that in the weak interference setting, although both the rates of Theorem \ref{thm:IAN2} and \cite[Thm. 3]{Hoydis:13} are computed assuming that intercell interference is treated as noise, the achievable average ergodic rates of Theorem \ref{thm:IAN2} are higher, with gains of $2.2$ bps/Hz and $0.45$ bps/Hz compared to  matched filtering and \ac{mmse} filtering, respectively, at high \acp{snr}, indicating that the \ac{se} of massive \ac{mimo} networks can be improved by allowing the \acp{bs} to perform joint decoding. 
We emphasize that an average ergodic rate gain of $0.45$ bps/Hz is translated into an overall ergodic rate gain of over $100$ bps/Hz in a cell with over $\Nusers = 220$ \acp{ut}. 
Additionally, the \ac{se} of treating interference as noise coincides with that of the optimized scheme, which settles with the known theoretical result that for the two-user Gaussian interference, treating interference as noise is optimal in the weak interference regime \cite[Ch. 6.4.3]{ElGamal:11}.  
Furthermore, as was also noted in the illustrative example in Subsection \ref{subsec:Example},  in high \acp{snr}, the performance of treating interference as noise is larger by a factor of approximately $\Ncells = 5$ compared to simultaneous decoding and time division. 

In the strong interference scenario, we observe in Fig. \ref{fig:NumEx2c}  that the optimized scheme as well as simultaneous decoding achieve an  average ergodic rate of $0.55$ bps/Hz,  while  separate decoding   results in negligible achievable rates, again, in agreement with the fact that simultaneous decoding is optimal in the strong interference regime for the two-user Gaussian interference channel, \cite[Ch. 6.4.2]{ElGamal:11}. Consequently, the fundamental limits of such channels are substantially higher than those achieved using standard  separate  linear decoding and treating interference as noise. 

For the moderate interference setting, none of the schemes 1-3 achieves the performance of the optimized scheme, and thus there is a clear gain in combining these schemes using the optimized scheme of Subsection \ref{subsec:Optimize}.  
This gain follows since in this case, the received signal at each \ac{bs} is impaired by notable intercell interference from some cells, and is hardly effected by the interference caused by other cells. Consequently, in this scenario, the fact that the optimized scheme allows treating the intercell interference caused by each cell differently is beneficial.
For the weak interference and strong interference settings, whose results are depicted in Figs. \ref{fig:NumEx2a} and \ref{fig:NumEx2c}, respectively, the optimized scheme does not utilize time-division, i.e., $\Nparts = 1$ and  $\{\myZeta_q\}_{q=1}^{\Nparts} = \{1\}$. However, for the moderate interference setting, for which some of the intercell interference is neither too weak nor too dominant, it is observed in Fig. \ref{fig:NumEx2b} that utilizing time-division is beneficial.
In particular, the optimized scheme here divides the transmission phase into $\Nparts = 2$ intervals. The first interval, which is utilized by $3$ cells, consists of $\myZeta_1 \approx 0.65$ of the transmission phase, while the remaining two cells utilize the rest  of the transmission phase.  
Using this assignment, in high \acp{snr},  the optimized scheme obtains a \ac{se} which is higher by $0.04$ bps/Hz compared to treating interference as noise when combining all three schemes, and by $0.018$ when combining only the decoding schemes 1-2, illustrating the benefit of combining time division.
We also note that for all schemes, the achievable rates hardly vary with \ac{snr} at high \acp{snr}, settling with the observation in \cite[Sec. IV]{Marzetta:10}.

Next, we numerically evaluate the dependence of the asymptotic \ac{se} on the number of pilot symbols. 
The purpose of this study is to check whether increasing $\Tpilots$, which increases the channel estimation accuracy at the cost of reducing the portion of the coherence interval used for data transmission, is beneficial in terms of \ac{se}. It is emphasized that increasing  $\Tpilots$ can also contribute to reducing the effect of pilot contamination by supporting different pilot reuse factors \cite{Jose:11}. However, to maintain consistency with the model used throughout the paper, in the following study we keep the pilot reuse factor to one, i.e., the same pilots are used in all the cells.
In Fig. \ref{fig:NumEx3a} we depict the \ac{se} versus the number of pilot symbols $\Tpilots$ at \ac{snr} of $0$ dB for the weak interference setting. 
Observing Fig.  \ref{fig:NumEx3a}, we note that, since the coherence duration is finite, for all the considered schemes, increasing the number of pilots  linearly decreases the \ac{se}. A similar behavior was observed with the moderate interference and strong interference settings. 
We also note that the ratios between the \acp{se} of the different schemes noted in Fig. \ref{fig:NumEx2a} for $\Tpilots = 100$, is approximately maintained also for larger values of $\Tpilots$.

\begin{figure}
 	\centering
 	\begin{minipage}{0.45\textwidth}
 		\centering
 		\scalebox{0.48}{\includegraphics{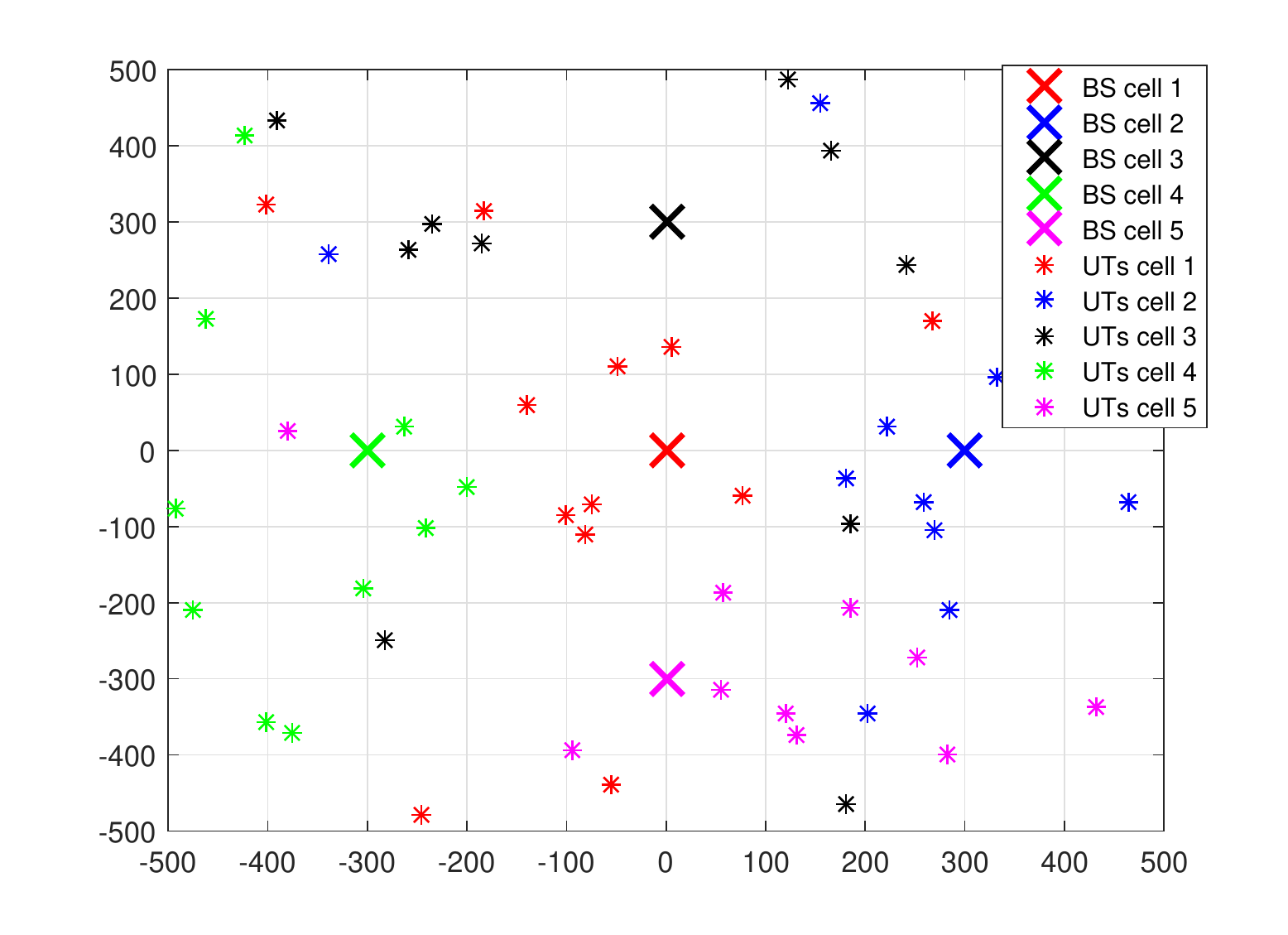}}
 		\vspace{-0.4cm}
 		\caption{Network layout with $\Nusers = 10$ and $\myProb = 0.8$.}% $\Nusers = 80$,  $\Nantennas = 800$.}
 		\label{fig:NumEx4a}		
 	\end{minipage}
 	$\quad$
 	\begin{minipage}{0.45\textwidth}
 		\centering
 		\scalebox{0.48}{\includegraphics{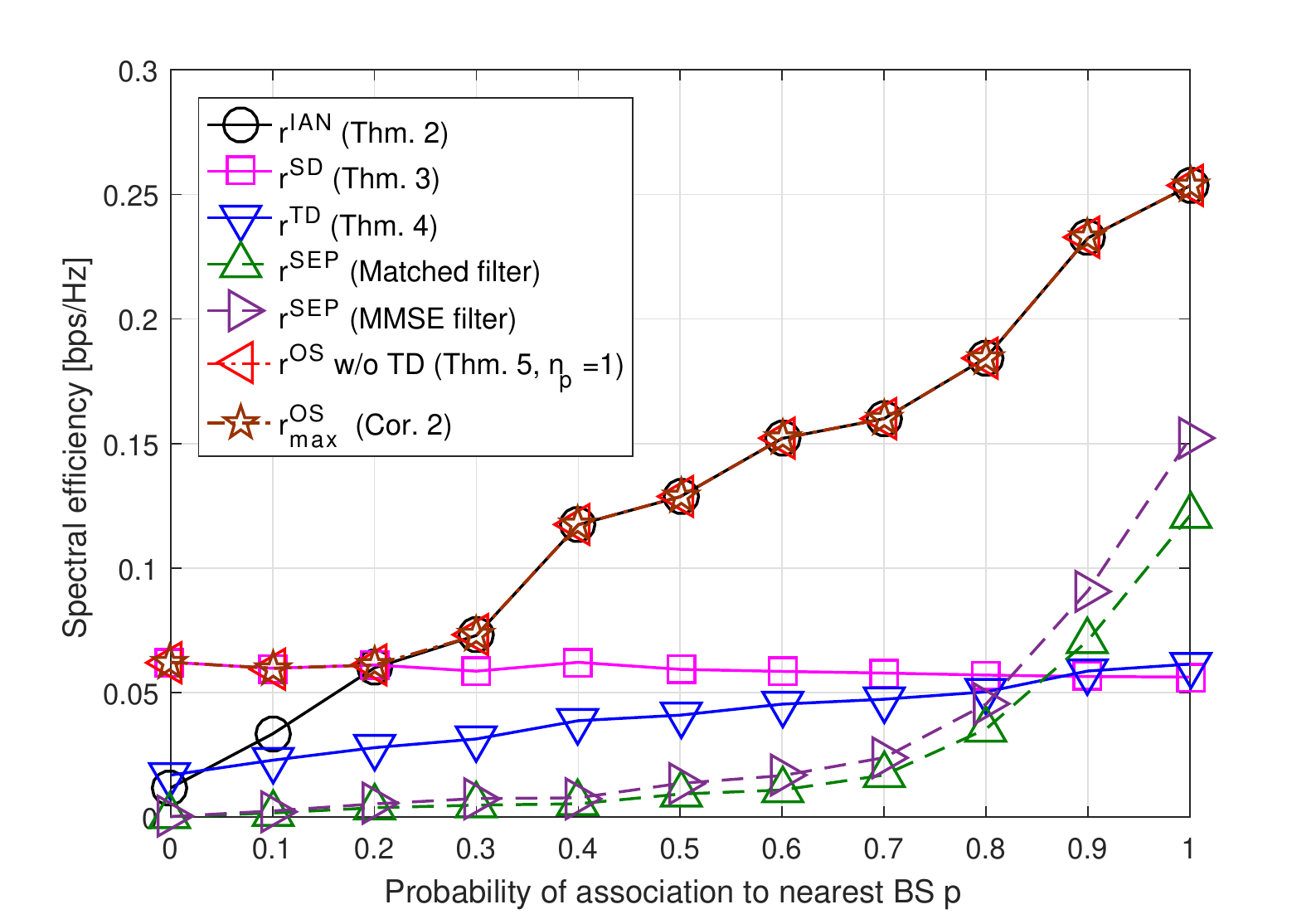}}
 		\vspace{-0.4cm}
 		\caption{\ac{se} versus probability of association to nearest \ac{bs}.}% $\Nusers = 80$,  $\Nantennas = 800$.}
 		\label{fig:NumEx4b}
 	\end{minipage}
 	\vspace{-0.3cm}
 \end{figure}

In our last simulations study, we numerically evaluate how the \ac{se} of each of the considered schemes depends on the level of the intercell interference in practical massive \ac{mimo} setups. To that aim, we consider an area of one square kilometer, in which  $\Ncells = 5$ are placed such the cell of index $k=1$ is located in the center of the grid, and the rest of the \acp{bs} are located at equally spaced points on a circle with radius of $300$ meters. 
Here, $C_{k,l,m}$ represents the distance from the $m$-th \ac{ut} of the $l$-th cell to the $k$-th \ac{bs}.
The location of each \ac{ut} is uniformly distributed over the considered area. Each \ac{ut} is associated to a \ac{bs} based on the following rule: For a fixed $\myProb \in [0,1]$, the \ac{ut} is assigned with probability $\myProb$ to the nearest \ac{bs}, and with equal probability of $\frac{1-\myProb}{4}$ to either of the other \acp{bs}. Such assignments can arise when the \ac{ut}-cell association rule accounts for additional objectives, aside from the standard reference signal received power, see, e.g., \cite{Bethanabhotla:16}.  
An illustration of a realization of such a network with $\Nusers = 10$ \acp{ut}  and  $\myProb = 0.8$ is depicted in Fig. \ref{fig:NumEx4a}. 
It is noted that as $\myProb$ increases, it is more likely that each \ac{ut} is associated with its nearest \ac{bs}, thus the intercell interference becomes less dominant. Consequently, by letting  $\myProb$ vary from $0$ to $1$, we are able to control the level of intercell interference in the network. 

In Fig. \ref{fig:NumEx4b} we depict the \acp{se} of the considered schemes versus $\myProb$ for \ac{snr} of $0$ dB. Observing Fig. \ref{fig:NumEx4b} we note that, as expected, for all values of $\myProb$, the optimized scheme of Subsection \ref{subsec:Optimize} achieves the best performance. In particular, for small values of $\myProb$, its performance coincides with that of simultaneous decoding, as the intercell interference is dominant. However, as $\myProb$ increases, the effect of intercell interference is reduced, and treating interference as noise becomes optimal. Furthermore, it is illustrated that the standard approach of separate linear decoding achieves poor \ac{se} for most intercell interference levels, and is able to provide reasonable performance only for $\myProb \ge 0.9$, namely, only when each \ac{ut} is associated with its nearest \ac{bs} with very high probability.

The results presented in this section demonstrate the potential benefits in terms of \ac{se} of properly acknowledging the nature of massive \ac{mimo} systems as interfering \acp{mac}.  
Furthermore, our results indicate  the fundamental performance limits of such channels, and how far the conventional approach for massive \ac{mimo} systems is from capturing these characteristics.

\label{txt:NewSimEnd}

The results presented in this section demonstrate the potential benefits in terms of \ac{se} of properly acknowledging the nature of massive \ac{mimo} systems as interfering \acp{mac}.  
Furthermore, our results indicate  the fundamental performance limits of such channels, and how far the conventional approach for massive \ac{mimo} systems is from capturing these characteristics.  

%----------------------------------------------------------------------------------------
%	CONCLUSIONS
%----------------------------------------------------------------------------------------
\vspace{-0.2cm}
\section{Conclusions}
\label{sec:Conclusions}
\vspace{-0.1cm}
In this paper we studied the \ac{se} of uplink massive \ac{mimo} systems when the \acp{bs} are allowed to jointly decode the received signals. We characterized the achievable average ergodic rates of three schemes for handling the intercell interference, in both the finite and asymptotic antenna regimes, and studied a method which combines these approaches for handling the intercell interference, aimed at maximizing the \ac{se}. 
Simulation results demonstrate the gains obtained by allowing the \acp{bs} to perform joint decoding, and indicate that in some scenarios, the standard approach of separate linear decoding fails to capture the fundamental performance limits of massive \ac{mimo} systems, especially when the interference is dominant.
The proposed analysis gives rise to a multitude of research paths, including the study of the \ac{se} with joint decoding under different system models, as well as the  analysis and the derivation of network decoding schemes in presence of additional design objectives.

%----------------------------------------------------------------------------------------
%	APPENDICES
%---------------------------------------------------------------------------------------- 
\begin{appendix}
\numberwithin{proposition}{subsection} 
\numberwithin{lemma}{subsection} 
\numberwithin{corollary}{subsection} 
\numberwithin{remark}{subsection} 
\numberwithin{equation}{subsection}	
%
%-----------------------------------
%	Proof of channel estimation lemma
%----------------------------------- 
\subsection{Proof of Lemma \ref{lem:ChEstLem}}
\label{app:Proof1} 
In order to obtain the \ac{mmse} estimate of $\Gmat_{k,l}$, we let $\Smat_p$ be the $\Tpilots \times \Tpilots$ unitary matrix (up to a fixed scaling constant) obtained from the full basis expansion of $\Smat$. 
Since $\Smat_p$ is deterministic and non-singular, it holds that the \ac{mmse} estimate satisfies
\vspace{-0.2cm}
\begin{align*}
\hat{\Gmat}_{k,l}  &= \E\left\{\Gmat_{k,l} | \myYmat_k, \{\Dmat_{k,l}  \} \right\}  
= \E\left\{\Gmat_{k,l} | \myYmat_k\Smat_p^H, \{\Dmat_{k,l}  \} \right\}  \notag \\
&\stackrel{(a)}{=} \E\left\{\Gmat_{k,l} | \myYmat_k\Smat^H, \{\Dmat_{k,l}  \} \right\}, 
\vspace{-0.2cm}
\end{align*}
where $(a)$ follows since, due the orthogonality of $\Smat_p$, the rows of $ \myYmat_k\Smat_p^H$ which do not belong to $\myYmat_k\Smat^H$ contain only noise which, given $ \{\Dmat_{k,l}\}$,  is independent of $\myYmat_k\Smat^H$, and $\{\Gmat_{k,l}\}$. In particular, $\myYmat_k\Smat^H$ is a sufficient statistics of $\Gmat_{k,l}$  from  $ \myYmat_k$ given $\{\Dmat_{k,l}  \} $ \cite[Ch. 2.9]{Cover:06}.
		
Next, we note that by \eqref{eqn:Channel_Training_MarForm} it holds that
$\myYmat_k\Smat^H = \Tpilots\sum\limits_{l=1}^{\Ncells}\Gmat_{k,l} + \myWmat_k\Smat^H$. Thus, given $ \{\Dmat_{k,l}\}$, the entries of  $\myYmat_k\Smat^H$ are mutually independent, and each entry of $\hat{\Gmat}_{k,l}$ can be independently estimated from its corresponding entry of $\myYmat_k\Smat^H $. Since, given $ \{\Dmat_{k,l}\}$, $\myYmat_k\Smat^H$ and $\Gmat_{k,l}$ are jointly Gaussian, the \ac{mmse} estimate of each entry is linear. Using the definition of  $\Bmat_{k,l}$ in \eqref{eqn:BmatDef}, it can be shown that $\hat{\Gmat}_{k,l}  $ is given by
  \begin{equation}
  \hat{\Gmat}_{k,l}  =   \myYmat_k  \mySmat^H \Tpilots^{-1} \Bmat_{k,l}, 
  \label{eqn:Proof1_4a}
  \end{equation}
thus proving \eqref{eqn:ChEstLem1}. 
Next, we study the statistical characterization of $\hat{\Gmat}_{k,l}$.    
Note that by \eqref{eqn:Proof1_4a}, 
\begin{align}
\hat{\Gmat}_{k,l} % = \myYmat_k  \mySmat^H \Tpilots^{-1} \Bmat_{k,l} 
&\stackrel{(a)}{=} \left(  \sum\limits_{l'=1}^{\Ncells}\Gmat_{k,l'} \mySmat + \myWmat_k\right)  \mySmat^H \Tpilots^{-1} \Bmat_{k,l}
 \notag \\
& 
\stackrel{(b)}{=} \left( \sum\limits_{l'=1}^{\Ncells} \Hmat_{k,l'}\Dmat_{k,l'}\right)\Bmat_{k,l} + \Tpilots^{-1}\myWmat_k \mySmat^H  \Bmat_{k,l}, 
\label{eqn:Proof1_5}
\end{align}
where $(a)$ follows from  the expression for $\myYmat_k$ in \eqref{eqn:Channel_Training_MarForm}, and $(b)$ follows since $\Gmat_{k,l} = \Hmat_{k,l}\Dmat_{k,l}$ and $\mySmat \mySmat^H = \Tpilots \myI_{\Nusers}$. 
%
%In the following we detail the statistical characterization of $\hat{\Gmat}_{k,l}$. % as the corresponding characterization of $\tilde{\Gmat}_{k,l}$ is obtained using similar arguments.
Since the entries of $\myWmat_k$ are i.i.d. zero-mean Gaussian \acp{rv} with variance $\SigW$,  the fact that  $\mySmat \mySmat^H = \Tpilots \myI_{\Nusers}$  implies that the entries of the matrix $\myWmat_k \mySmat^H$   are i.i.d. zero-mean Gaussian \acp{rv} with variance $\SigW \Tpilots$. 
Consequently,  for a given realization $\{\Dmat_{k,l} = \DmatRel_{k,l}\}$ with diagonal coefficients   $\{\dcoeff_{k,l,m} = \dcoeffRel_{k,l,m}\}$, we have that the entries of the diagonal matrix $\Bmat_{k,l}$ are given by the deterministic values $\left( \Bmat_{k,l} \right)_{m,m}  =  \frac{\Tpilots \dcoeffRel_{k,l,m}^2}{\SigW + \Tpilots\sum\limits_{l'=1}^{\Ncells} \dcoeffRel_{k,l',m}^2 } \triangleq \left( \BmatRel_{k,l} \right)_{m,m}$. It thus follows from \eqref{eqn:Proof1_5} that the entries of $\hat{\Gmat}_{k,l}$ are zero-mean mutually independent Gaussian \acp{rv} with variance 
% ---- NIR SINGLE COLUMN VERSION START ----	 
\ifsingle
%\vspace{-0.2cm}
%\begin{equation*}
$\E\Big\{\big|\big( \hat{\Gmat}_{k,l}\big)_{m_1,m_2}   \big|^2 \Big|  \{\Dmat_{k,l} \!=\! \DmatRel_{k,l}\} \Big\}
%= \left(  \sum\limits_{l'=1}^{\Ncells} \dcoeffRel_{k,l',m_2}^2 + \SigW  \Tpilots^{-1} \right)  \Bigg( \frac{\Tpilots \dcoeffRel_{k,l,m_2}^2}{\SigW + \Tpilots\sum\limits_{l'=1}^{\Ncells} \dcoeffRel_{k,l',m_2}^2 }\Bigg)^2  
%\notag \\
%&\qquad \qquad \qquad \qquad \qquad 
= \frac{\Tpilots \dcoeffRel_{k,l,m_2}^2}{\SigW + \Tpilots\sum\limits_{l'=1}^{\Ncells} \dcoeffRel_{k,l',m_2}^2 } \dcoeffRel_{k,l,m_2}^2 
%\notag \\
%&
= \left( \BmatRel_{k,l} \right)_{m_2,m_2}  \left( \DmatRel_{k,l} \right)_{m_2,m_2}^2$. 
%\vspace{-0.2cm}
%\end{equation*}
\else
\begin{align}
&\E\left\{\left|\left( \hat{\Gmat}_{k,l}\right)_{m_1,m_2}   \right|^2 |  \{\Dmat_{k,l} = \DmatRel_{k,l}\}_{l \in \NcellsSet} \right\} \notag \\
&= \left(  \sum\limits_{l'=1}^{\Ncells} \dcoeffRel_{k,l',m_2}^2 + \SigW  \Tpilots^{-1} \right)  \left( \frac{\Tpilots \dcoeffRel_{k,l,m_2}^2}{\SigW + \Tpilots\sum\limits_{l'=1}^{\Ncells} \dcoeffRel_{k,l',m_2}^2 }\right)^2  
\notag \\
&= \frac{\Tpilots \dcoeffRel_{k,l,m_2}^2}{\SigW + \Tpilots\sum\limits_{l'=1}^{\Ncells} \dcoeffRel_{k,l',m_2}^2 } \dcoeffRel_{k,l',m_2}^2 \notag \\
&= \left( \BmatRel_{k,l} \right)_{m_2,m_2}  \left( \DmatRel_{k,l} \right)_{m_2,m_2}^2. 
\end{align}
\fi %------------------------------------------
Accordingly, the conditional distribution of any set of entries from $\hat{\Gmat}_{k,l}$ given  $\{\Dmat_{k,l} = \DmatRel_{k,l}\}$ is identical to the conditional distribution of the corresponding set of entries from  
$\UHmat\Bmat_{k,l}^{1/2}\Dmat_{k,l}$ given  $\{\Dmat_{k,l} = \DmatRel_{k,l}\}$, recalling that $\UHmat$ is a zero-mean Gaussian random matrix with i.i.d. unit variance entries independent of  $\{\Dmat_{k,l}\}_{l \in \NcellsSet}$. It thus follows from the law of total probability  \cite[Ch. 8.2]{Papoulis:91} that 
  $\hat{\Gmat}_{k,l} \Dist \UHmat\Bmat_{k,l}^{1/2}\Dmat_{k,l}$. 
  The proof that $\tilde{\Gmat}_{k,l} \Dist \UHmat\left( \myI_{\Nusers} - \Bmat_{k,l}\right)^{1/2}\Dmat_{k,l}$ is obtained using similar arguments and is thus omitted for brevity.
%  TODO NIR - finish the characterization of $\tilde{\Gmat}_{k,l}$
\qed

%-----------------------------------
%	Proof of IAN proposition
%----------------------------------- 
\subsection{Proof of Proposition \ref{lem:IAN1}}
\label{app:Proof2}   
To prove the proposition, we first formulate the achievable ergodic sum-rate for the $k$-th \ac{bs} using the covariance matrix of $\myV\IAN{k}[i]$ conditioned on $\hat{\Gmat}_{k,k}$ and  $\{\Dmat_{k,l}\}$, denoted $\CovMat{\myV\IAN{k}  | \hat{\Gmat}_{k,k}, \{\Dmat_{k,l}\}}$. 
Then, we obtain an achievable  ergodic sum-rate  which depends on the covariance matrix of ${\myV}\IAN{k}[i]$ conditioned only on   $\{\Dmat_{k,l}\}$, denoted $\CovMat{{\myV}\IAN{k}  |   \{\Dmat_{k,l}\}}$.
Finally, we prove that the resulting achievable ergodic sum-rate yields the achievable average ergodic rate  given in \eqref{eqn:IAN1}.

Let us first consider the achievable ergodic sum-rate of the \ac{mac} whose input-output relationship is given in \eqref{eqn:IANModel2} for a fixed $k\in\NcellsSet$.
During data transmission, the $k$-th \ac{bs} knows the attenuation coefficients $\{\Dmat_{k,l} \}$  and the estimated channel, $\hat{\Gmat}_{k,k}$. Conditioned on these \acp{rv}, the estimation error $\tilde{\Gmat}_{k,k}$ is zero-mean, since, by the law of total expectation \cite[Ch. 7.4]{Papoulis:91}, $\E\left\{\tilde{\Gmat}_{k,k} | \hat{\Gmat}_{k,k} , \{\Dmat_{k,l} \}\right\}  = \E\left\{\E\left\{{\Gmat}_{k,k}  | \hat{\Gmat}_{k,k} , \myYmat_k, \{\Dmat_{k,l} \right\} |  \hat{\Gmat}_{k,k} , \{\Dmat_{k,l} \}\right\} - \hat{\Gmat}_{k,k}$, and thus, 
% ---- NIR SINGLE COLUMN VERSION START ----	 
\ifsingle
\vspace{-0.2cm}
\begin{equation}
\E\left\{\tilde{\Gmat}_{k,k} | \hat{\Gmat}_{k,k} , \{\Dmat_{k,l} \}\right\} 
%&= \E\left\{{\Gmat}_{k,k} - \hat{\Gmat}_{k,k}  | \hat{\Gmat}_{k,k} , \{\Dmat_{k,l} \}\right\} \notag \\
%&\stackrel{(a)}{=}  \E\left\{\E\left\{{\Gmat}_{k,k}  | \hat{\Gmat}_{k,k} , \myYmat_k, \{\Dmat_{k,l} \right\} |  \hat{\Gmat}_{k,k} , \{\Dmat_{k,l} \}\right\} - \hat{\Gmat}_{k,k} \notag \\
%&
\stackrel{(a)}{=} \E\left\{\hat{\Gmat}_{k,k} |  \hat{\Gmat}_{k,k} , \{\Dmat_{k,l} \}\right\}- \hat{\Gmat}_{k,k} 
= 0,
\vspace{-0.2cm}
\label{eqn:uncorrelated1}
\end{equation}
\else
\begin{align}
&\E\left\{\tilde{\Gmat}_{k,k} | \hat{\Gmat}_{k,k} , \{\Dmat_{k,l} \}\right\} 
= \E\left\{{\Gmat}_{k,k} \!- \! \hat{\Gmat}_{k,k}  | \hat{\Gmat}_{k,k} , \{\Dmat_{k,l} \}\right\} \notag \\
&\stackrel{(a)}{=}  \E\left\{\E\left\{{\Gmat}_{k,k}  | \hat{\Gmat}_{k,k} , \myYmat_k, \{\Dmat_{k,l} \}\right\} |  \hat{\Gmat}_{k,k} , \{\Dmat_{k,l} \}\right\} \! -\! \hat{\Gmat}_{k,k} \notag \\
&\stackrel{(b)}{=} \E\left\{\hat{\Gmat}_{k,k} |  \hat{\Gmat}_{k,k} , \{\Dmat_{k,l} \}\right\}- \hat{\Gmat}_{k,k} 
= 0,
\label{eqn:uncorrelated1}
\end{align}
\fi %--------------------------
where $(a)$ %follows from the law of total expectation \cite[Ch. 7.4]{Papoulis:91}, and $(b)$ 
follows since $\hat{\Gmat}_{k,k}$ is the \ac{mmse} estimate of ${\Gmat}_{k,k}$ given $\myYmat_k$, $\{\Dmat_{k,l} \}$, %thus $\E\big\{{\Gmat}_{k,k}  | \hat{\Gmat}_{k,k} , \myYmat_k, \{\Dmat_{k,l} \}\big\} %\! =\! \E\big\{{\Gmat}_{k,k}  | \myYmat_k, \{\Dmat_{k,l} \}\big\} 
%\! =\! \hat{\Gmat}_{k,k}$.
%
Consequently, 
the equivalent noise $\myV\IAN{k}[i]$ is orthogonal to $\myX_k[i]$, thus \eqref{eqn:IANModel2} represents a \ac{mac} with an additive uncorrelated noise $\myV\IAN{k}[i]$ and a known channel matrix $\hat{\Gmat}_{k,k}$. Since the worst-case additive uncorrelated noise distribution is Gaussian  
\cite[Thm. 1]{Hassibi:03}\footnote{Although \cite{Hassibi:03} considered \ac{ptp} \ac{mimo} channels, for a fixed input distribution, the achievable sum-rate of a \ac{mac} is equal to the achievable rate of a \ac{ptp} \ac{mimo} channel with the same input-output relationship. Hence, \cite[Thm. 1]{Hassibi:03} applies also to \acp{mac}.},  
 the achievable ergodic sum-rate of the \ac{mac} \eqref{eqn:IANModel2} with Gaussian $\myV\IAN{k}[i]$ is also achievable with any other distribution of $\myV\IAN{k}[i]$. 

By letting the codelength span a sufficiently large number of realizations of  $\{\Dmat_{k,l}\} $  and $\{\Hmat_{k,l}\} $, noting that the \ac{bs} knows the channel attenuations and the \ac{mmse} estimate of the channel, the following ergodic sum-rate is achievable for the \ac{mac} \eqref{eqn:IANModel2} \cite[Ch. 23.5]{ElGamal:11}: 
\begin{align}
 \sum\limits_{m=1}^{\Nusers}\!\R{k,m} 
&=  I\left(\myX_k ; \myY_k | \hat{\Gmat}_{k,k}, \{\Dmat_{k,l}\}   \right) \notag \\
&\stackrel{(a)}{\ge} \E \left\{\log \left|\myI_{\Nantennas} \!+\! \hat{\Gmat}_{k,k}\hat{\Gmat}_{k,k}^H \CovMat{\myV\IAN{k}  | \hat{\Gmat}_{k,k}, \{\Dmat_{k,l}\}}^{-1} \right|   \right\} \notag \\
&= \E \left\{\log \left|\CovMat{\myV\IAN{k}  | \hat{\Gmat}_{k,k}, \{\Dmat_{k,l}\}} \!+\! \hat{\Gmat}_{k,k}\hat{\Gmat}_{k,k}^H  \right|   \right\} \notag \\ 
&- \E \left\{\log \left|\CovMat{\myV\IAN{k}  | \hat{\Gmat}_{k,k}, \{\Dmat_{k,l}\}} \right|   \right\},
%\vspace{-0.2cm}
\label{eqn:MI1}
\end{align}
where $(a)$ follows by computing the mutual information for Gaussian additive uncorrelated noise  $\myV\IAN{k}[i]$   \cite[Ch. 9.1]{ElGamal:11}, as the worst-case additive noise is Gaussian. 

Next, we explicitly express the matrix $\CovMat{\myV\IAN{k}  | \hat{\Gmat}_{k,k}, \{\Dmat_{k,l}\}}$. Note that  from \eqref{eqn:BmatDef} and  \eqref{eqn:ChEstLem1},  $\hat{\Gmat}_{k,l} = \hat{\Gmat}_{k,k} \Dmat_{k,k}^{-2}\Dmat_{k,l}^2$, and therefore, 
%\begin{equation}
%\vspace{-0.2cm}
${\myV}\IAN{k}[i] 
%&= \sum\limits_{l=1}^{\Ncells}\tilde{\Gmat}_{k,l}\myX_l[i] + \sum\limits_{l=1, l\ne k}^{\Ncells}\hat{\Gmat}_{k,l}\myX_l[i] + \myW_k[i] \notag \\
%&\stackrel{(a)}{=} 
=\sum\limits_{l=1}^{\Ncells}\tilde{\Gmat}_{k,l}\myX_l[i] + \hat{\Gmat}_{k,k}\Dmat_{k,k}^{-2}\sum\limits_{l=1, l\ne k}^{\Ncells}\Dmat_{k,l}^{2}\myX_l[i] + \myW_k[i]$.
%\vspace{-0.2cm}
%\label{eqn:VIANdef1}
%\end{equation}
%where $(a)$ follows from \eqref{eqn:BmatDef} and  \eqref{eqn:ChEstLem1} as $\hat{\Gmat}_{k,l} = \hat{\Gmat}_{k,k} \Dmat_{k,k}^{-2}\Dmat_{k,l}^2$.  
As $\{\tilde{\Gmat}_{k,l}\}$ and $\hat{\Gmat}_{k,k}$ are jointly Gaussian and uncorrelated given  $\{\Dmat_{k,l}\}$, then,  
% $\CovMat{\myV\IAN{k}  | \hat{\Gmat}_{k,k}, \{\Dmat_{k,l}\}}$ can be written as
$\CovMat{{\myV}\IAN{k}  |  \hat{\Gmat}_{k,k},  \{\Dmat_{k,l}\}} =  \sum\limits_{l=1 }^{\Ncells} \E\left\{ \tilde{\Gmat}_{k,l} \tilde{\Gmat}_{k,l}^H  |  \{\Dmat_{k,l}\}  \right\}  + \hat{\Gmat}_{k,k}\Dmat_{k,k}^{-4}\sum\limits_{l=1, l\ne k}^{\Ncells}\Dmat_{k,l}^{4}  \hat{\Gmat}_{k,k}^H  +\SigW \myI_{\Nantennas}$, which yields
 \vspace{-0.3cm}
\begin{align}
\CovMat{{\myV}\IAN{k}  |  \hat{\Gmat}_{k,k},  \{\Dmat_{k,l}\}}
%&=   \sum\limits_{l=1 }^{\Ncells} \E\left\{ \tilde{\Gmat}_{k,l} \tilde{\Gmat}_{k,l}^H  | \hat{\Gmat}_{k,k}, \{\Dmat_{k,l}\}  \right\}  + \hat{\Gmat}_{k,k}\Dmat_{k,k}^{-4}\sum\limits_{l=1, l\ne k}^{\Ncells}\Dmat_{k,l}^{4}  \hat{\Gmat}_{k,k}^H  +\SigW \myI_{\Nantennas} \notag \\
%&=    \sum\limits_{l=1 }^{\Ncells} \E\left\{ \tilde{\Gmat}_{k,l} \tilde{\Gmat}_{k,l}^H  |  \{\Dmat_{k,l}\}  \right\}  + \hat{\Gmat}_{k,k}\Dmat_{k,k}^{-4}\sum\limits_{l=1, l\ne k}^{\Ncells}\Dmat_{k,l}^{4}  \hat{\Gmat}_{k,k}^H  +\SigW \myI_{\Nantennas} \notag \\
%&
&\stackrel{(a)}{=} \myT_k \cdot\myI_{\Nantennas} \notag \\
& + \hat{\Gmat}_{k,k}\Dmat_{k,k}^{-4}\sum\limits_{l=1, l\ne k}^{\Ncells}\Dmat_{k,l}^{4}  \hat{\Gmat}_{k,k}^H,
\vspace{-0.2cm}
\label{eqn:CovMat11}
\end{align}
where $(a)$ follows  from Lemma \ref{lem:ChEstLem}, as
%\vspace{-0.2cm}
%\begin{align}
% \sum\limits_{l=1 }^{\Ncells} \E\left\{ \tilde{\Gmat}_{k,l} \tilde{\Gmat}_{k,l}^H  |  \{\Dmat_{k,l}\}  \right\} \!+\!\SigW \myI_{\Nantennas}  &=
%\sum\limits_{l=1 }^{\Ncells} \E\left\{\UHmat\left( \myI_{\Nusers} \!-\! \Bmat_{k,l} \right) \Dmat_{k,l}^2 \UHmat^H   | \{\Dmat_{k,l}\} \right\} \!+\!\SigW \myI_{\Nantennas} \notag \\
%&\stackrel{(b)}{=}  \sum\limits_{l=1 }^{\Ncells} {\rm Tr}\left(\left( \myI_{\Nusers} \!-\! \Bmat_{k,l} \right) \Dmat_{k,l}^2 \right) \myI_{\Nantennas}\! +\!\SigW \myI_{\Nantennas} = \myT_k \cdot\myI_{\Nantennas} ,
%\vspace{-0.2cm}
%\label{eqn:CovMat12}
%\end{align}
%and $(b)$ follows since
 for any  $\myMat{Q}$,
$\E\{\UHmat   \myMat{Q} \UHmat ^H  \} = {\rm Tr}\left( \myMat{Q}  \right) \myI_{\Nantennas}$ \cite[Sec. III-B]{Soysal:10a}.

Substituting \eqref{eqn:CovMat11} into \eqref{eqn:MI1}, recalling that  $\hat{\Gmat}_{k,k} \Dist \UHmat\Bmat_{k,k}^{1/2}\Dmat_{k,k}$, where  $ \Bmat_{k,k} $ and $ \Dmat_{k,k}$ are diagonal matrices with strictly positive diagonal entries, results in  
\begin{align}
\sum\limits_{m=1}^{\Nusers}\!\R{k,m} 
 &\ge \E \left\{\log \left|\myI_{\Nantennas} + \UHmat \AmatUp\UHmat^H  \right|  \right\} \notag \\ 
 &- \E \left\{\log \left|\myI_{\Nantennas} + \UHmat \AmatDn\UHmat^H  \right|  \right\},
 %\vspace{-0.2cm}
\label{eqn:MI1s}
\end{align}
where
% $(a)$ follows since by  Lemma \ref{lem:ChEstLem}, $\hat{\Gmat}_{k,k} \Dist \UHmat\Bmat_{k,k}^{1/2}\Dmat_{k,k}$, where  $ \Bmat_{k,k} $ and $ \Dmat_{k,k}$ are diagonal matrices with strictly positive diagonal entries, and from the definition of
 $\AmatUp, \AmatDn$ are defined in \eqref{eqn:AggMatDefIAN}.
This proves  that $\RIAN{\Nantennas}$ given in \eqref{eqn:IAN1} is  achievable. % when treating intercell interference as noise. 
\qed

%-----------------------------------
%	Proof of IAN theorem
%----------------------------------- 
\subsection{Proof of Theorem \ref{thm:IAN2}}
\label{app:Proof3} 
We prove the theorem by applying Theorem \ref{thm:MarPast1} to characterize  \eqref{eqn:IAN1} in the limit $\Nantennas \rightarrow \infty$ with $\frac{\Nusers}{\Nantennas}=\AntRatio$. To that aim, we first show that the conditions of Theorem \ref{thm:MarPast1} are satisfied, and then we apply Theorem \ref{thm:MarPast1} to obtain \eqref{eqn:IAN2}. We now explicitly derive  $\mathop{\lim}\limits_{{\Nantennas \rightarrow \infty}}\E \left\{\frac{1}{\Nantennas}\log \left|\myI_{\Nantennas} + \UHmat \AmatUp\UHmat^H  \right|  \right\}$; the derivation of this limit with $\AmatUp$ replaced by $\AmatDn$ is similar and thus omitted for brevity.

 As the entries of $\UHmat$ are i.i.d. unit variance \acp{rv} independent of $\AmatUp$, the  matrix  $\UHmat \AmatUp\UHmat^H = \frac{1}{\Nantennas} \UHmat \big(\Nantennas \cdot \AmatUp\big) \UHmat^H$ satisfies the conditions of  Theorem \ref{thm:MarPast1} when the empirical eigenvalue distribution of $\Nantennas \cdot \AmatUp$ converges to a non-random limit almost surely.  
Since $\Nantennas \cdot \AmatUp$ is a diagonal matrix, its eigenvalues are given by its diagonal entries 
% \vspace{-0.2cm}
%\begin{equation}
$\left( \Nantennas \!\cdot\! \AmatUp\right)_{m,m} 
%=  \frac{\bcoeff_{k,k,m}\dcoeff_{k,k,m}^{-2} \sum\limits_{l=1}^{\Ncells}\dcoeff_{k,l,m}^{4}}{\frac{1}{\Nantennas}\sum\limits_{l=1}^{\Ncells} \sum\limits_{m'=1}^{\Nusers}\left(1 \!- \!\bcoeff_{k,l,m'} \right)\dcoeff_{k,l,m'}^2\! + \!  \frac{1}{\Nantennas} \SigW } \notag \\
%&\qquad
=  \frac{\bcoeff_{k,k,m}\dcoeff_{k,k,m}^{-2} \sum\limits_{l=1}^{\Ncells}\dcoeff_{k,l,m}^{4}}{\AntRatio  \sum\limits_{l=1}^{\Ncells} \!\big(\! \frac{1}{\Nusers}\!\sum\limits_{m'=1}^{\Nusers}\!\left(1\! -\! \bcoeff_{k,l,m'} \right)\dcoeff_{k,l,m'}^2 \!\big)\! +\!   \frac{1}{\Nantennas} \SigW }$,
% \vspace{-0.2cm}
%\label{eqn:EmpAsym1}
%\end{equation}
for $m \in \NusersSet$. From \eqref{eqn:BmatDef}, it follows that for any $k,l \in \NcellsSet$ the \acp{rv} $\big\{\left(1 - \bcoeff_{k,l,m'} \right)\dcoeff_{k,l,m'}^2 \big\}_{m' \in \NusersSet}$ are i.i.d., and thus, by  the strong law of large numbers \cite[Ch. 2.4]{Durret:10}, $\frac{1}{\Nusers}\sum\limits_{m'=1}^{\Nusers}\left(1 - \bcoeff_{k,l,m'} \right)\dcoeff_{k,l,m'}^2 $ converges almost surely to $\E\{\left(1 - \bcoeff_{k,l,1} \right)\dcoeff_{k,l,1}^2 \}$. Consequently, it follows from   \cite[Ch. 20.6]{Cramer:70} that for sufficiently large $\Nantennas$ with fixed $\frac{\Nusers}{\Nantennas} = \AntRatio$, the distribution of the eigenvalues of $\Nantennas \cdot \AmatUp$ approaches the distribution of the set of i.i.d. \acp{rv} $\Big\{\frac{\bcoeff_{k,k,m}\dcoeff_{k,k,m}^{-2} \sum\limits_{l=1}^{\Ncells}\dcoeff_{k,l,m}^{4}}{\AntRatio  \sum\limits_{l=1}^{\Ncells} \E\{\left(1 - \bcoeff_{k,l,1} \right)\dcoeff_{k,l,1}^2 \} } \Big\}_{m \in \NusersSet}$. It therefore follows from \cite[Thm. 2.4.7]{Durret:10} that the empirical \ac{cdf} of the  eigenvalues of $\Nantennas \cdot \AmatUp$ converges almost surely to the non-random \ac{cdf} of the random variable $\AvarUp$ defined in \eqref{eqn:AIANkDef1}, and that the random matrix $\UHmat \AmatUp\UHmat^H$ satisfies the conditions of  Theorem \ref{thm:MarPast1}.  
Consequently, in the massive \ac{mimo} regime, the achievable average ergodic rate in \eqref{eqn:IAN1} can be written as in \eqref{eqn:IAN2}. 
\qed

%-----------------------------------
%	Proof of SD proposition
%----------------------------------- 
\subsection{Proof of Proposition \ref{lem:SD1}}
\label{app:Proof4}   
When each \ac{bs} decodes the messages of all \acp{ut} in the network, the input-output relationship \eqref{eqn:SDModel2} represents a set of $\Ncells$ \acp{mac} with $\Ncells \cdot \Nusers$ transmitters. Thus, 
letting the codelength span a sufficiently large number of realizations of $\{\Dmat_{k,l}\} $  and  $\{\Hmat_{k,l}\} $, as the \ac{bs} knows the attenuation coefficients and the \ac{mmse} channel estimate,
every sum-rate which satisfies  
\begin{equation}
\label{eqn:Proof4a} 
\sum\limits_{l=1}^{\Ncells}\sum\limits_{m=1}^{\Nusers} \! \R{l,m} \!\le\! I \!\left( \myX_1, \myX_2, \ldots, \myX_{\Ncells} ; {\myY}_k | \hat{\Gmat}_{k,k}, \{\Dmat_{k,l}\}  \right),    
\end{equation}
$\forall k \in \NcellsSet, $
is an achievable ergodic sum-rate \cite[Ch. 23.5]{ElGamal:11}. 

Let $\CovMat{{\myV}\SD{k}  |   \{\Dmat_{k,l}\}}$  and $\CovMat{{\myV}\SD{k}  |  \hat{\Gmat}_{k,k}, \{\Dmat_{k,l}\}}$ be the covariance matrices of ${\myV}\SD{k}[i]$ conditioned on    $\{\Dmat_{k,l}\} $ and on  $\hat{\Gmat}_{k,k},\{\Dmat_{k,l}\} $, respectively. 
Repeating the arguments in \eqref{eqn:uncorrelated1}, we have that the equivalent noise $\myV\SD{k}[i] = \sum\limits_{l=1}^{\Ncells}\tilde{\Gmat}_{k,l}\myX_l[i]+ \myW_k[i]$ is orthogonal to $\myX_l[i]$ for every $k,l \in \NcellsSet$.   
Since the worst-case additive uncorrelated noise distribution is Gaussian \cite[Thm. 1]{Hassibi:03}, by computing the mutual information \eqref{eqn:Proof4a} with Gaussian  $\myV\SD{k}[i]$ we have that \cite[Ch. 9.1]{ElGamal:11} $I \left( \myX_1, \myX_2, \ldots, \myX_{\Ncells} ; {\myY}_k | \hat{\Gmat}_{k,k}, \{\Dmat_{k,l}\}  \right) 
\ge \E \left\{\log \left|\myI_{\Nantennas} + \hat{\Gmat}_{k,k}\Dmat_{k,k}^{-4} \sum\limits_{l=1}^{\Ncells}\Dmat_{k,l}^{4}  \hat{\Gmat}_{k,k}^H \CovMat{{\myV}\SD{k}  | \hat{\Gmat}_{k,k}\{\Dmat_{k,l}\}}^{-1} \right|   \right\}$. 
As  $\{\Dmat_{k,l}\}$ are diagonal matrices with strictly positive diagonal entries, and since  given $ \{\Dmat_{k,l}\} $, each \ac{mmse} estimate $\hat{\Gmat}_{k,l}$ is jointly Gaussian and uncorrelated with the estimation error $\tilde{\Gmat}_{k,l}$, it follows that ${\myV}\SD{k}[i]$ is independent of $\hat{\Gmat}_{k,k}$ given $ \{\Dmat_{k,l}\} $, and thus
% ---- NIR SINGLE COLUMN VERSION START ----	 
\ifsingle
\vspace{-0.2cm}
\begin{equation}
\hspace{-0.2cm}
I \left( \myX_1, \myX_2, \ldots, \myX_{\Ncells} ; {\myY}_k | \hat{\Gmat}_{k,k}, \{\Dmat_{k,l}\}  \right) 
\!\ge\! 
%\E \left\{\log \left|\myI_{\Nantennas} + \hat{\Gmat}_{k,k}\Dmat_{k,k}^{-4} \sum\limits_{l=1}^{\Ncells}\Dmat_{k,l}^{4}  \hat{\Gmat}_{k,k}^H \CovMat{{\myV}\SD{k}  | \hat{\Gmat}_{k,k}\{\Dmat_{k,l}\}}^{-1} \right|   \right\}
%\notag \\
%%&\quad\ge \E \left\{\log \left|\myI_{\Nantennas} + \hat{\Gmat}_{k,k}\Dmat_{k,k}^{-2} \sum\limits_{l=1}^{\Ncells}\Dmat_{k,l}^{2}   \left( \hat{\Gmat}_{k,k}\Dmat_{k,k}^{-2} \sum\limits_{l=1}^{\Ncells}\Dmat_{k,l}^{2}\right) ^H \CovMat{{\myV}\SD{k}  |  \hat{\Gmat}_{k,k},\{\Dmat_{k,l}\}}^{-1} \right|   \right\} \notag \\
%&\qquad \qquad \qquad \qquad \qquad \qquad \quad \stackrel{(a)}{=} 
\E \left\{\log \left|\myI_{\Nantennas} \!+\! \hat{\Gmat}_{k,k}\Dmat_{k,k}^{-4} \sum\limits_{l=1}^{\Ncells}\Dmat_{k,l}^{4}  \hat{\Gmat}_{k,k}^H \CovMat{{\myV}\SD{k}  | \{\Dmat_{k,l}\}}^{-1} \right|   \right\}.
\vspace{-0.2cm}
\label{eqn:Proof4b}
\end{equation}
\else
\begin{align}
&I \left( \myX_1, \myX_2, \ldots, \myX_{\Ncells} ; {\myY}_k | \hat{\Gmat}_{k,k}, \{\Dmat_{k,l}\}  \right) \notag \\
&\ge\! \E \bigg\{\!\log\! \Big|\myI_{\Nantennas} \!+\! \hat{\Gmat}_{k,k}\Dmat_{k,k}^{-2} \sum\limits_{l=1}^{\Ncells}\Dmat_{k,l}^{2}   \left( \hat{\Gmat}_{k,k}\Dmat_{k,k}^{-2} \sum\limits_{l=1}^{\Ncells}\Dmat_{k,l}^{2}\right) ^H \notag \\
&\qquad \qquad \qquad \qquad\qquad\qquad\qquad\times \CovMat{{\myV}\SD{k}  |  \hat{\Gmat}_{k,k},\{\Dmat_{k,l}\}}^{-1} \Big|   \bigg\} \notag \\
&\stackrel{(a)}{=} \!\E \bigg\{\!\log \!\Big|\myI_{\Nantennas} \!+\! \hat{\Gmat}_{k,k}\Dmat_{k,k}^{-4} \sum\limits_{l=1}^{\Ncells}\Dmat_{k,l}^{4}  \hat{\Gmat}_{k,k}^H \CovMat{{\myV}\SD{k}  | \{\Dmat_{k,l}\}}^{-1} \Big|   \bigg\},
\label{eqn:Proof4b}
\end{align}
\fi % ------------------------------------- 

Next,  repeating the arguments used in \eqref{eqn:CovMat11} to compute  $\CovMat{\tilde{\myV}\IAN{k}  |   \{\Dmat_{k,l}\}}$, we have that 
%\vspace{-0.2cm}
%\begin{equation}
$\CovMat{{\myV}\SD{k}  | \{\Dmat_{k,l}\}} 
=  \sum\limits_{l=1 }^{\Ncells} {\rm Tr}\left(\left( \myI_{\Nusers}\! -\! \Bmat_{k,l} \right) \Dmat_{k,l}^2 \right) \myI_{\Nantennas} \!+\!\SigW \myI_{\Nantennas} %\notag \\
= \myT_k^{-1} \cdot ^{} \myI_{\Nantennas}$. 
%\vspace{-0.2cm}
%\label{eqn:CovMat13}
%\end{equation}
Consequently, from Lemma \ref{lem:ChEstLem} and \eqref{eqn:AggMatDefIAN}, we have that 
%\vspace{-0.2cm}
%\begin{equation}
$\hat{\Gmat}_{k,k}\Dmat_{k,k}^{-4} \sum\limits_{l=1}^{\Ncells}\Dmat_{k,l}^{4}  \hat{\Gmat}_{k,k}^H  \CovMat{{\myV}\SD{k}  | \{\Dmat_{k,l}\}}^{-1}
\Dist 
%\myT_k^{-1} \cdot \UHmat  \Bmat_{k,k} \Dmat_{k,k}^{-2} \sum\limits_{l=1}^{\Ncells}\Dmat_{k,l}^{4}  \UHmat^H 
%\stackrel{(a)}{=}  
 \UHmat \AmatUp\UHmat^H$,
%\vspace{-0.2cm}
%\label{eqn:Proof4c}
%\end{equation}
%where $(a)$ follows from \eqref{eqn:AggMatDefIAN1}.  
Combining this with  \eqref{eqn:Proof4b} yields
% ---- NIR SINGLE COLUMN VERSION START ----	 
\ifsingle
\vspace{-0.2cm}
\begin{equation}
I \left( \myX_1, \myX_2, \ldots, \myX_{\Ncells} ; {\myY}_k | \hat{\Gmat}_{k,k}, \{\Dmat_{k,l}\}  \right) 
\ge \E \left\{\log \left|\myI_{\Nantennas} +  \UHmat \AmatUp\UHmat^H   \right|\right\}.
\label{eqn:Proof4d}
\vspace{-0.2cm}
\end{equation}
\else
\begin{align}
&I \left( \myX_1, \myX_2, \ldots, \myX_{\Ncells} ; {\myY}_k | \hat{\Gmat}_{k,k}, \{\Dmat_{k,l}\}  \right) \notag \\
& \qquad \quad
\ge \E \left\{\log \left|\myI_{\Nantennas} +  \UHmat \AmatUp\UHmat^H   \right|\right\}.
\label{eqn:Proof4d}
\end{align}
\fi % -------------------------------
It thus follows from \eqref{eqn:Proof4a} and \eqref{eqn:Proof4d} that $\mathop{\min}\limits_{k \in \NcellsSet} \E \left\{\log \left|\myI_{\Nantennas} +  \UHmat \AmatUp\UHmat^H \right|  \right\}$ is an achievable ergodic sum-rate for the \acp{mac} given by \eqref{eqn:SDModel2}, and thus, $\RSD{\Nantennas}$ given in \eqref{eqn:SD1} is an achievable average ergodic rate  when the \acp{bs} decode the  intercell interference, proving the proposition.
\qed

%-----------------------------------
%	Proof of TD proposition
%-----------------------------------
\vspace{-0.25cm}
\subsection{Proof of Proposition \ref{lem:TD1}}
\label{app:Proof6}  
\vspace{-0.15cm}
When the intercell interference is eliminated using time-division, the input-output relationship \eqref{eqn:TDModel2} represents a set of $\Ncells$ \acp{mac}, each with $\Nusers$ transmitters. Thus, 
letting the codelength span a sufficiently large number of realizations of the attenuation coefficients $\{\Dmat_{k,l}\} $  and channel matrices $\{\Hmat_{k,l}\} $, as the \ac{bs} knows the attenuation coefficients and the \ac{mmse} channel estimate, the following ergodic sum-rate is achievable for the $k$-th \ac{mac} \eqref{eqn:TDModel2}, $ k \in \NcellsSet$  \cite[Ch. 23.5]{ElGamal:11}:
\vspace{-0.2cm}
\begin{equation}
\label{eqn:Proof6a}
\hspace{-0.4cm}
\sum\limits_{m=1}^{\Nusers} \! \R{k,m} \!\le\! I \!\left(  \myX_k ; {\myY}_k | \hat{\Gmat}_{k,k}, \{\Dmat_{k,l}\}  \right).  
\vspace{-0.2cm}
\end{equation}

Let $\CovMat{{\myV}\TD{k}  |   \{\Dmat_{k,l}\}}$  be the covariance matrices of ${\myV}\TD{k}[i]$ conditioned on    $\{\Dmat_{k,l}\}$. Note that $\myV\TD{k}  = \tilde{\Gmat}_{k,k} \myZeta_k^{-\frac{1}{2}} \myX_k[i] + \myW_k[i]$ is independent of the \ac{mmse} estimate $\hat{\Gmat}_{k,k}$ given $\{\Dmat_{k,l}\}$, and orthogonal to $\myX_k[i]$ for every $k \in \NcellsSet$.   
Since the worst-case additive uncorrelated noise distribution is Gaussian \cite[Thm. 1]{Hassibi:03}, by computing  \eqref{eqn:Proof6a} with Gaussian  $\myV\TD{k}[i]$ we have that \cite[Ch. 9.1]{ElGamal:11}
% ---- NIR SINGLE COLUMN VERSION START ----	 
\ifsingle	
%\vspace{-0.1cm}
%\begin{equation}
$I\left(\myX_k ; {\myY}_k | \hat{\Gmat}_{k,k}, \{\Dmat_{k,l}\}   \right)
\ge   \E \left\{\log \left|\myI_{\Nantennas} + \hat{\Gmat}_{k,k}\hat{\Gmat}_{k,k}^H \CovMat{{\myV}\TD{k}  | \{\Dmat_{k,l}\}}^{-1} \right|   \right\}$. 
%\vspace{-0.1cm}
%\label{eqn:aid6d}
%\end{equation}
\else
\begin{align}
&I\left(\myX_k ; {\myY}_k | \hat{\Gmat}_{k,k}, \{\Dmat_{k,l}\}   \right)\notag \\
&\qquad\ge \E \left\{\log \left|\myI_{\Nantennas} + \myZeta_k^{-1}\hat{\Gmat}_{k,k}\hat{\Gmat}_{k,k}^H \CovMat{{\myV}\TD{k}  | \{\Dmat_{k,l}\}}^{-1} \right|   \right\}. 
\label{eqn:aid6d}
\end{align}
\fi %-------------------------------
Next,  repeating the arguments used in \eqref{eqn:CovMat11} to compute  $\CovMat{\tilde{\myV}\IAN{k}  |   \{\Dmat_{k,l}\}}$, we have that 
%\vspace{-0.2cm}
%\begin{equation}
$\CovMat{{\myV}\TD{k}  | \{\Dmat_{k,l}\}} 
=    \myZeta_k^{-1}{\rm Tr}\left(\left( \myI_{\Nusers}\! -\! \Bmat_{k,k} \right) \Dmat_{k,k}^2 \right) \myI_{\Nantennas} \!+\!\SigW \myI_{\Nantennas}$. 
%\label{eqn:CovMat32}
%\vspace{-0.2cm}
%\end{equation}
Thus, from Lemma \ref{lem:ChEstLem} and \eqref{eqn:AggMatDefTD},
$\myZeta_k^{-1}\hat{\Gmat}_{k,k}\hat{\Gmat}_{k,k}^H \CovMat{{\myV}\TD{k}  | \{\Dmat_{k,l}\}}^{-1} 
\Dist \UHmat\AmatTD\!\left( \myZeta_k\right) \UHmat^H$. 
From  \eqref{eqn:Proof6a}, we  have that $\E \left\{\log \left|\myI_{\Nantennas} \!+\!  \UHmat \AmatTD\!\left( \myZeta_k\right) \UHmat^H   \right|\right\}$ is an achievable ergodic sum-rate for the  \ac{mac} whose input-output relationship is given in \eqref{eqn:TDModel2}. As each \ac{mac} uses only $\myZeta_k$ of the data transmission phase, the  \ac{se} is given in \eqref{eqn:TD1}, proving the proposition.
\qed

%-----------------------------------
%	Proof of OS theorem
%----------------------------------- 
\subsection{Proof of Proposition \ref{pro:Example}}
\label{app:Proof9} 
To prove the proposition, we first express the \acp{rv}  $\AvarUp$, $\AvarDn$, and $\AvarTD$, for the considered setup, and the corresponding \acp{se} $\RIAN{}$, $\RSD{}$, and $\RTD{}$. Then, we use these expressions to characterize the relationships between the asymptotic \acp{se} when $\Ymax \ll \Xmin$ and when $\Xmax \ll \Ymin$. 

First, we note that for the considered setup, the \acp{rv} $\bcoeff_{k,l,m}$ defined in \eqref{eqn:BmatDef} are distributed via $\bcoeff_{k,l,m} \Dist \frac{\Tpilots\Xrv}{\SigW + \Tpilots(\Xrv + \Yrv)}$ for $k = l$ and $\bcoeff_{k,l,m} \Dist \frac{\Tpilots\Yrv}{\SigW + \Tpilots(\Xrv + \Yrv)}$ for $k \ne l$, for each $m \in \NusersSet$. 
Consequently, by defining $\mxy \triangleq \E \big\{\frac{\SigW + \Tpilots \cdot \Xrv \dot \Yrv}{\SigW + \Tpilots ( \Xrv + \Yrv)} \big\} \underset{\SigW \to 0}{=} \E \big\{\frac{\Xrv \cdot \Yrv}{\Xrv + \Yrv} \big\}$, for each $k = 1,2$, $\AvarUp$ defined in \eqref{eqn:AIANkDef1} satisfies 
$\AvarUp = \frac{\bcoeff_{k,k,1}\dcoeff_{k,k,1}^{-2} (\dcoeff_{k,1,1}^{4} + \dcoeff_{k,2,1}^{4}) } {\AntRatio  \left(  \E\{\left(1 - \bcoeff_{k,1,1} \right)\dcoeff_{k,1,1}^2 \} +  \E\{\left(1 - \bcoeff_{k,2,1} \right)\dcoeff_{k,2,1}^2 \} \right) } $, and thus
\vspace{-0.2cm}
\begin{align}
\AvarUp
\Dist &\frac{\Tpilots}{\AntRatio \cdot \mxy}  \frac{\Xrv^2 + \Yrv^2}{\SigW + \Tpilots(\Xrv + \Yrv)} \notag \\
&\stackrel{(a)}{=} \frac{1}{\AntRatio \cdot \mxy}  \frac{\Xrv^2 + \Yrv^2}{\Xrv + \Yrv},
\vspace{-0.2cm}
\label{eqn:Avar1}
\end{align}
where $(a)$ follows since $\SigW \rightarrow 0$. Similarly, the \acp{rv} $\AvarDn$ and $\AvarTD$   satisfy
\vspace{-0.2cm}
\begin{equation}
\AvarDn \Dist   \frac{1}{\AntRatio \cdot \mxy}  \frac{\Yrv^2}{\Xrv + \Yrv}, \quad  
\AvarTD \Dist   \frac{1}{\AntRatio \cdot \mxy}  \frac{\Xrv^2}{\Xrv + \Yrv},
\label{eqn:Avar3}
\end{equation}
for each $k  = 1,2$.
It follows \eqref{eqn:Avar1}-\eqref{eqn:Avar3} that the distribution of the \acp{rv} $\AvarUp$, $\AvarDn$, and $\AvarTD$ does not depend on $k$, and thus the asymptotic \acp{se} in \eqref{eqn:IAN2}, \eqref{eqn:SD2}, and \eqref{eqn:TDOpt2}, satisfy for any $k = 1,2$
\begin{subequations}
\label{eqn:TD2a}
\begin{eqnarray} 
\RIAN{} &=&\frac{\Tdata}{\Tcoh \cdot \AntRatio}\!\left(  \MPFunc\left(\AvarUp, \AntRatio \right) \!- \! \MPFunc\left(\AvarDn, \AntRatio \right)\right);  \\
\RSD{} &=&\frac{\Tdata}{2\Tcoh \cdot \AntRatio} \MPFunc\left(\AvarUp, \AntRatio \right);  \\
\RTD{} &=&\frac{\Tdata}{2\Tcoh \cdot \AntRatio} \MPFunc\left(\AvarTD, \AntRatio \right). 
\vspace{-0.2cm}
\end{eqnarray}
\end{subequations}

To characterize the relationship between $\RSD{}$ and $\RIAN{}$,  we use the following lemma:
\begin{lemma}
	\label{lem:Aid1}
	For  an \ac{rv} $A$ satisfying $\Pr\left( 0 \le A < a_{\max}\right) =1$, 
	if  $\AntRatio \cdot \E \left\{ \frac{ \myEta \cdot A}{1 + \myEta \cdot A} \right\} < \frac{1}{2}$, where $\myEta \in (0,1]$ is given in Theorem \ref{thm:MarPast1}, then 
		$\MPFunc\left( A, \AntRatio\right) \le \AntRatio\cdot \log(1 +a_{\max})  + \log e \cdot \left(\AntRatio  \cdot  \frac{  a_{\max}}{1 +  a_{\max}}  \right)^2$.  
\end{lemma}
\begin{IEEEproof}
Note   that  $\myEta\! =\! 1\! -\! \AntRatio \cdot \E \left\{ \frac{ \myEta \cdot A}{1 + \myEta \cdot A} \right\}$. For $\AntRatio \cdot \E \left\{ \frac{ \myEta \cdot A}{1 + \myEta \cdot A} \right\} < \frac{1}{2}$, plugging this into \eqref{eqn:MarPast1} yields
		\vspace{-0.2cm}
	\begin{align}
	&\MPFunc\left( A, \AntRatio\right) 
	%	&= \AntRatio\cdot \E \left\{\log(1 +\myEta \cdot A) \right\} +  \log e \cdot \left(\myEta - 1 - \log_e \myEta \right) \notag \\
	= \AntRatio\cdot \E \left\{\log(1 \! + \!\myEta \cdot A) \right\} \! + \!  \log e \notag\\ 
	&\qquad \cdot \left( - \AntRatio \cdot \E \left\{ \frac{ \myEta \cdot A}{1 \! + \! \myEta \cdot A} \right\} - \log_e \left( 1 - \AntRatio \cdot \E \left\{ \frac{ \myEta \cdot A}{1 \! + \! \myEta \cdot A} \right\}\right)  \right) \notag \\
	&\stackrel{(a)}{\le} \AntRatio\cdot \E \left\{\log(1 \! + \!\myEta \cdot A) \right\} \! + \! \log e \cdot \left(\!\AntRatio  \cdot \E \left\{ \frac{  \myEta \cdot A}{1 \! + \!  \myEta \cdot A} \right\} \right)^2\!\! ,
%	\notag \\
%	&\stackrel{(b)}{\le} \AntRatio\cdot \log(1 \! + \!a_{\max})  \! + \! \log e \cdot \left(\AntRatio  \cdot  \frac{  a_{\max}}{1 \! + \!  a_{\max}}  \right)^2,
	\label{eqn:Aid1b1}
		\vspace{-0.2cm}
	\end{align}
	where $(a)$ follows since for $\alpha \in [0, \frac{1}{2}]$, $-\alpha - \alpha^2 \le \log_e (1 -\alpha)$. As $\frac{   \alpha}{1 +   \alpha}$ and $\log(1+\alpha)$ are monotonically non-decreasing and $\eta \cdot A \le a_{\max}$, \eqref{eqn:Aid1b1} proves the lemma. 
\end{IEEEproof}

We can now prove that when $\Ymax \ll \Xmin$, $\RIAN{} \approx 2\RSD{}$. 
From \eqref{eqn:TD2a} we have that
%	\vspace{-0.2cm}
%\begin{equation}
%\label{eqn:WeakIntEq1}
$\RIAN{} = 2\cdot \RSD{} - \frac{\Tdata}{\Tcoh \cdot \AntRatio} \MPFunc\left(\AvarDn, \AntRatio \right)$.
%	\vspace{-0.2cm}
%\end{equation}
Next, we prove that $\AvarDn$ satisfies the conditions of Lemma \ref{lem:Aid1}.
note that $\AvarDn \le \frac{1}{\AntRatio \cdot \mxy} \frac{\Ymax^2}{\Ymax + \Xmin}\triangleq a_{\max}$ with probability one, and thus $\AntRatio \cdot \E \left\{ \frac{ \myEta \cdot \AvarDn}{1 + \myEta \cdot \AvarDn} \right\} \le \AntRatio \cdot  \frac{ a_{\max}}{1 + a_{\max}}$. Furthermore, since $\Yrv \ll \Xrv$ with probability one,  we have that $\mxy \approx \E\{\Yrv\}$, and thus  $a_{\max} \approx \frac{\Ymax}{\AntRatio \cdot \E\{\Yrv\}} \frac{\Ymax}{ \Xmin}$. Consequently, since $\Ymax \ll \Xmin$ then $a_{\max} \approx 0$, and thus  $\AntRatio \cdot \E \left\{ \frac{ \myEta \cdot \AvarDn}{1 + \myEta \cdot \AvarDn} \right\} < \frac{1}{2}$. 
Thus, $\AvarDn$ satisfies the conditions of Lemma \ref{lem:Aid1}, and therefore, 
%	\vspace{-0.1cm}
%\begin{equation}
$\MPFunc\left( \AvarDn, \AntRatio\right) 
\le \AntRatio\cdot \log(1 +a_{\max})  + \log e \cdot \left(\AntRatio  \cdot  \frac{  a_{\max}}{1 +  a_{\max}}  \right)^2  \stackrel{(a)}{\approx} 0$, 
%	\vspace{-0.1cm}
%\label{eqn:WeakIntEq2}
%\end{equation}
where $(a)$ follows since  $a_{\max}$ tends to zero. Consequently,  $\RIAN{} \approx 2\RSD{}$.

Lastly, we consider the case in which $\Xmax \ll \Ymax$.
 Here, we have that $\Xrv \ll \Yrv$ with probability one. In this case it follows from \eqref{eqn:Avar1} and \eqref{eqn:Avar3} that for any $k = 1,2$, the distribution of the \acp{rv} $\AvarUp$ and $\AvarDn$ approaches the distribution of the \ac{rv} $ \frac{\Yrv}{\AntRatio \cdot \mxy}$. Consequently, by \eqref{eqn:TD2a}, we have that $\RIAN{} \approx 0$.  Similarly, the distribution of $\AvarTD$ approaches the distribution of the \ac{rv} $ \frac{1}{\AntRatio \cdot \mxy}\frac{\Xrv^2}{\Yrv}$. Consequently, by   \eqref{eqn:TD2a}, 
%\begin{equation}
$\RSD{} \approx\frac{\Tdata}{2\Tcoh \cdot \AntRatio} \MPFunc\left(\frac{\Yrv}{\AntRatio \cdot \mxy}, \AntRatio \right)$, and
$\RTD{} \approx\frac{\Tdata}{2\Tcoh \cdot \AntRatio} \MPFunc\left(\frac{1}{\AntRatio \cdot \mxy}\frac{\Xrv^2}{\Yrv}. \AntRatio \right)$, 
%\label{eqn:TD2b}
%\end{equation}
Now, by considering the same network in which the \acp{ut} of cell $k=1$ are allocated to to cell $k=2$ and vice versa, we have that  the \ac{se} of treating interference as noise, which is  strictly positive, is given by  $\frac{\Tdata}{\Tcoh \cdot \AntRatio} \big(\MPFunc\left(\frac{\Yrv}{\AntRatio \cdot \mxy}, \AntRatio \right) \!-\! \MPFunc\left(\frac{1}{\AntRatio \cdot \mxy}\frac{\Xrv^2}{\Yrv}. \AntRatio \right)\big) > 0$. 
Thus,  $\RSD{} > \RTD{}$.  % proving the proposition.
\qed

%-----------------------------------
%	Proof of OS theorem
%----------------------------------- 
\subsection{Proof of Theorem \ref{thm:OS2}}
\label{app:Proof8} 
To prove the theorem, we first obtain the \ac{se} in the finite antenna regime, and then we let $\Nantennas$ tend to infinity and use Theorem \ref{thm:MarPast1} to obtain \eqref{eqn:OS2}. 
From the representation in \eqref{eqn:OSModel2}, by treating $\myV\OS{k}$ as noise and decoding the interference $\{\myX_l\}_{l \in \Clust, l \ne k}$, we have that $\myY_k$ is the output of a \ac{mac} with $|\Clust| \cdot \Nusers$ transmitters. 
Consequently, 
 by repeating the arguments in the proofs of Propositions \ref{lem:IAN1}-\ref{lem:TD1}, we have that for each cluster $\Clust$, every sum-rate with satisfies that 
\begin{equation}
\label{eqn:Proof8a}
\sum\limits_{l\in \Clust}\sum\limits_{m=1}^{\Nusers} \! \R{l,m} \!\le\! I \!\left( \{\myX_l\}_{l \in \Clust} ; {\myY}_k | \hat{\Gmat}_{k,k}, \{\Dmat_{k,l}\}  \right),  
\end{equation}
 $\forall k \in \Clust$, 
is an achievable ergodic sum-rate \cite[Ch. 23.5]{ElGamal:11}. 

\begin{figure*}[!t]
	\normalsize
	\begin{align}
	& I \!\left( \{\myX_l\}_{l \in \Clust} ; {\myY}_k | \hat{\Gmat}_{k,k}, \{\Dmat_{k,l}\}  \right) 
	\ge \E \bigg\{\log \bigg|\myI_{\Nantennas}\! +\! \myZeta_q^{-1}\hat{\Gmat}_{k,k}\Dmat_{k,k}^{-4} \sum\limits_{l\in \Clust}\Dmat_{k,l}^{4}  \hat{\Gmat}_{k,k} ^H \CovMat{{\myV}\OS{k}  |  \hat{\Gmat}_{k,k},\{\Dmat_{k,l}\}}^{-1} \bigg|   \bigg\} \notag \\
	& = \E \bigg\{\log \bigg|\CovMat{{\myV}\OS{k}  |  \hat{\Gmat}_{k,k},\{\Dmat_{k,l}\}} \!+\! \myZeta_q^{-1}\hat{\Gmat}_{k,k}\Dmat_{k,k}^{-4} \sum\limits_{l\in \Clust}\Dmat_{k,l}^{4}  \hat{\Gmat}_{k,k} ^H  \bigg|   \bigg\} \!-\! \E \left\{\log \left|\CovMat{{\myV}\OS{k}  |  \hat{\Gmat}_{k,k},\{\Dmat_{k,l}\}}  \right|   \right\}.  
	\label{eqn:Proof8b}
	\end{align}
	\begin{align}
	&\CovMat{\myV\OS{k} |\hat{\Gmat}_{k,k}, \{\Dmat_{k,l}\}} 
	= \myZeta_q^{-1}\sum\limits_{l \in \Part{q} }\E\left\{ \tilde{\Gmat}_{k,l} \tilde{\Gmat}_{k,l}^H  | \hat{\Gmat}_{k,k}, \{\Dmat_{k,l}\}  \right\}  + \myZeta_q^{-1}\hat{\Gmat}_{k,k}\Dmat_{k,k}^{-4}\sum\limits_{l\in \NegClust}\Dmat_{k,l}^{4}  \hat{\Gmat}_{k,k}^H  +\SigW \myI_{\Nantennas} \notag \\
	&\quad \stackrel{(a)}{=} \myZeta_q^{-1}\bigg(  \sum\limits_{l\in \Part{q} } {\rm Tr}\big(( \myI_{\Nusers} - \Bmat_{k,l} ) \Dmat_{k,l}^2 \big)  +\myZeta_q\SigW\bigg) \myI_{\Nantennas}  + \myZeta_q^{-1}\hat{\Gmat}_{k,k}\Dmat_{k,k}^{-4}\sum\limits_{l\in \NegClust}\Dmat_{k,l}^{4}  \hat{\Gmat}_{k,k}^H,
	\label{eqn:Proof8c} 
	\end{align}
	% IEEE uses as a separator
	\hrulefill
	% The spacer can be tweaked to stop underfull vboxes.
	\vspace*{4pt}
\end{figure*}

Let $\CovMat{\myV\OS{k} |\hat{\Gmat}_{k,k}, \{\Dmat_{k,l}\}}$ be the covariance matrix of the equivalent noise $\myV\OS{k}$ given $\hat{\Gmat}_{k,k}, \{\Dmat_{k,l}\}$. By worst-case additive uncorrelated noise arguments, recalling that $\myY_k[i] = \myZeta_q^{-\frac{1}{2}} \hat{\Gmat}_{k,k}{\Dmat}_{k,k}^{-2}\sum\limits_{l\in \Clust}{\Dmat}_{k,l}^2\myX_l[i] + \myV\OS{k}[i]$, we have that the conditional mutual information is bounded as in \eqref{eqn:Proof8b} .
Next, we note that the covariance matrix $\CovMat{\myV\OS{k} |\hat{\Gmat}_{k,k}, \{\Dmat_{k,l}\}}$ can be written as in \eqref{eqn:Proof8c}, 
where $(a)$ follows from Lemma \ref{lem:ChEstLem} and since  for any  $\myMat{Q}$,
$\E\{\UHmat   \myMat{Q} \UHmat ^H  \} = {\rm Tr}\left( \myMat{Q}  \right) \myI_{\Nantennas}$ \cite[Sec. III-B]{Soysal:10a}.
Thus, by defining  $\tilde{T}_k \triangleq	 \sum\limits_{l\in \Part{q} } {\rm Tr}\big(( \myI_{\Nusers} - \Bmat_{k,l} ) \Dmat_{k,l}^2 \big)  +\myZeta_q\cdot\SigW$, 
 $	\TAmatUp \triangleq \tilde{T}_k^{-1} \Bmat_{k,k}\Dmat_{k,k}^{-2}.\sum\limits_{l\in \Part{q}}\Dmat_{k,l}^{4}$, and 
 ${\TAmatDn} \triangleq \tilde{T}_k^{-1} \Bmat_{k,k}\Dmat_{k,k}^{-2}.\sum\limits_{l\in \NegClust}\Dmat_{k,l}^{4}$, 
and substituting \eqref{eqn:Proof8c} into \eqref{eqn:Proof8b}, 
%\begin{equation}
%\hspace{-0.2cm}
$I \!\left( \{\myX_l\}_{l \in \Clust} ; {\myY}_k | \hat{\Gmat}_{k,k}, \{\Dmat_{k,l}\}  \right)  
\ge \E \left\{\log \left|\myI_{\Nantennas}\! +\! \UHmat {\TAmatUp} \UHmat^H  \right|  \right\} 
\!-\! \E \left\{\log \left|\myI_{\Nantennas} \!+\! \UHmat {\TAmatDn} \UHmat^H  \right|  \right\}$.
%\label{eqn:Proof8d}
%\end{equation}
Combining this with \eqref{eqn:Proof8a} implies that 
\begin{align*}
\sum\limits_{l\in \Clust}\sum\limits_{m=1}^{\Nusers} \! \R{l,m} = \mathop{\min}\limits_{k \in \Clust} \Big(&\E \big\{\log \big|\myI_{\Nantennas} + \UHmat {\TAmatUp} \UHmat^H  \big|  \big\} \notag \\
&- \E \big\{\log \big|\myI_{\Nantennas} + \UHmat {\TAmatDn} \UHmat^H  \big|  \big\}\Big)
\end{align*}
 is an achievable ergodic sum-rate. Consequently, as each \ac{mac} uses only $\myZeta_q$ of the data transmission phase, then 
% following average ergodic rate is achievable: 
\begin{align*}
\ROS{\Nantennas} \triangleq 
 \frac{\Tdata}{\Tcoh}\!\cdot\!\frac{1}{\Ncells  \cdot \Nantennas}  \sum\limits_{q =1}^{\Nparts} \myZeta_q  \sum\limits_{s=1}^{\Nclusts}&\mathop{\min}\limits_{k \in \Clust}\Big(\E \left\{\log \left|\myI_{\Nantennas}\! +\! \UHmat {\TAmatUp} \UHmat^H  \right|  \right\} \notag \\
 &- \E \left\{\log \left|\myI_{\Nantennas} + \UHmat {\TAmatDn} \UHmat^H  \right|  \right\}\Big), 
\end{align*} 
is achievable. 
It can be shown by repeating the arguments in the proof of Theorem \ref{thm:IAN2} that the random matrices $ \UHmat \big(\Nantennas \cdot{\TAmatUp}\big) \UHmat^H$ and $\UHmat \big(\Nantennas \cdot{\TAmatDn}\big) \UHmat^H$ satisfy the conditions of Theorem \ref{thm:MarPast1}, and thus, for $\Nantennas \rightarrow \infty$, $\ROS{\Nantennas}$ equals the right hand side of \eqref{eqn:OS2}, proving the theorem. 
\qed

\end{appendix} 

%----------------------------------------------------------------------------------------
%	BIBLIOGRAPHY
%---------------------------------------------------------------------------------------- 

\end{document}